\documentclass[12pt]{article}

\usepackage{color}
\usepackage[applemac]{inputenc}
\usepackage{hyperref}

\usepackage[sc,small]{caption}

\usepackage{amsmath,amssymb}

\usepackage{graphicx}

\numberwithin{equation}{section}

\DeclareRobustCommand{\SkipTocEntry}[4]{}
\newcommand{\be}{\begin{eqnarray}}
\newcommand{\ee}{\end{eqnarray}}
\newcommand{\tht}{\vartheta}

\newcommand{\N}{{\mathcal N}}

\newcommand{\beqn}{\begin{eqnarray}}
\newcommand{\eeqn}{\end{eqnarray}}
\newcommand{\tr}{{\rm tr}}

\newcommand{\q}{u}

\newcommand{\V}{\mathcal{V}}

\newcommand{\buzz}{superpotential de-sequestering }

\newcommand{\slsr}{\ensuremath{SU(2)_L \times SU(2)_R} }
\newcommand{\ab}{\ensuremath{\tr( A^i B^j)} }
\DeclareMathSymbol{\C}{\mathbin}{AMSb}{"43}
\newcommand{\ba}{ \left( \begin{array}{c c} }
\newcommand{\ea}{ \end{array} \right) }

\newcommand{\X}{X}
\newcommand{\x}{x}

\newcommand{\KAB}{\ensuremath{K^{A \bar{B}}}}
\newcommand{\bb}{\ensuremath{\bar{b}}}

\newcommand{\Ms}{\ensuremath{M^2_{Pl}}}
\newcommand{\M}{\ensuremath{M_{Pl}}}

\newcommand{\Zs}{\ensuremath{Z_{\star} }}
\newcommand{\axs}{\ensuremath{\vline_{Z_0} }}
\newcommand{\axst}{\ensuremath{\vline_{Z_{\star}} }}
\newcommand{\mgrs}{\ensuremath{m^2_{3/2}}}
\newcommand{\Kcd}{\ensuremath{K^{c \bar{d}}}}
\newcommand{\dt}{\ensuremath{\delta_{MNP}}}
\newcommand{\dts}{\ensuremath{\delta_{\{MN\}P}}}

\begin{document}
\thispagestyle{empty}

\ \\
\vspace{1cm}
\begin{center}
{\bf {\Large Sequestering in String Compactifications}}\\ \ \\ \ \\
Marcus Berg,${}^1$ David Marsh,${}^2$ Liam McAllister,${}^2$ and Enrico Pajer${}^2$\\
\vspace{1cm}
{\sl ${}^1$Oskar Klein Center for Cosmoparticle Physics and
Department of Physics,\\ Stockholm University, Albanova SE-106 91 Stockholm, Sweden \\
\vskip 4pt
${}^2$Department of Physics, Cornell University, Ithaca, NY 14853 USA }
\end{center}
\vspace{1cm}
\hrule \ \\
\noindent {\bf Abstract} \\
We study the mediation of supersymmetry breaking in string compactifications whose moduli are stabilized by nonperturbative effects. We begin with a critical review of arguments for sequestering in supergravity and in string theory. We then show that geometric isolation, even in a highly warped space, is insufficient to achieve sequestering: in type IIB compactifications, nonperturbative superpotentials involving the K\"ahler moduli introduce cross-couplings between well-separated visible and hidden sectors. The scale of the resulting soft terms depends on the moduli stabilization scenario. In the Large Volume Scenario, nonperturbative superpotential contributions to the soft trilinear $A$ terms can introduce significant flavor violation, while in KKLT compactifications their effects are negligible. In both cases, the contributions to the $\mu$ and $B\mu$ parameters cannot be ignored in general. We conclude that sequestered supersymmetry breaking is possible in nonperturbatively-stabilized compactifications only if a mechanism in addition to bulk locality suppresses superpotential cross-couplings.

\bigskip
\hrule
\vfill \today
\newpage

\tableofcontents

\newpage

\section{Introduction}

The stabilization of the electroweak scale is one of the most significant open questions in theoretical physics. Low-energy supersymmetry provides an  elegant solution, but supersymmetry must be broken, and the experimental signatures are principally governed by the supersymmetry-breaking soft terms.  In the most plausible scenarios, supersymmetry is broken in a hidden sector and this breaking is mediated to the visible sector by some form of interaction.  The structure of the resulting soft terms is largely controlled by the nature of the mediating interaction, motivating efforts to study the mediation of supersymmetry breaking without making reference to the details of the hidden sector.

Naturalness suggests that the visible and hidden sectors should be coupled by non-renormalizable operators induced by integrating out new interactions near the Planck scale.  In the celebrated gravity mediation scenario, these couplings provide the leading interaction between the two sectors, and give rise to soft scalar masses of order the gravitino mass $m_{3/2}$.
Unfortunately, extremely little is known about Planck-scale interactions, yet some  detailed properties of these interactions --- at least, as encoded in the structure of the effective theory below the Planck scale --- are required in order to make predictions.  For instance, strong bounds on flavor violation force the fermion and sfermion mass matrices to be diagonal in the same basis, to high accuracy.  It has proved difficult to justify symmetries of a
Planck-scale theory that can enforce such a flavor structure.  This supersymmetric  flavor problem is a serious obstacle to successful phenomenology in high-scale mediation.

The flavor problem in gravity-mediated supersymmetry breaking could be ameliorated if the soft masses in the visible sector were parametrically suppressed compared to $m_{3/2}$.  In this case it is possible for scalar mass contributions from some other mediation mechanism, e.g. anomaly mediation, to give rise to visible sector masses with acceptable flavor structure.  In such a situation, one says that the source of supersymmetry breaking is {\it{sequestered}} from the visible sector \cite{RS}.

Sequestering amounts to a suppression of the soft terms compared to the natural level induced by `generic' Planck-suppressed operators coupling the hidden and visible sectors.
This state of affairs is unnatural unless it is enforced by a symmetry or other structure in the Planck-scale theory, e.g. extradimensional locality, and the success or failure of sequestering depends very sensitively on Planck-scale interactions.  This strongly motivates studying sequestering in string theory, where such contributions can in principle be computed.

Randall and Sundrum originally proposed that sequestering could arise as a result of geometric separation in an internal space \cite{RS}: locality in the extra dimensions, where only gravity was assumed to propagate, severely restricted the form of the lower dimensional effective theory.   At first sight the extradimensional construction in \cite{RS} appears amenable to a realization in string theory,  but on closer examination the mechanism of spatial separation does not manifestly extend to string compactifications with moduli: light moduli could easily mediate interactions of gravitational strength, while sequestering requires far feebler interactions.  Indeed, Anisimov et al. \cite{Anisimov:2002az} have argued that for precisely this reason, sequestering is difficult to obtain in certain classes of string compactifications, while Kachru et al. \cite{Kachru:2006em} observed that even stabilized moduli are very generally too light to decouple in the manner required.

Nevertheless, it was shown in \cite{KMS} that sequestering is natural in certain highly-warped string compactifications: this is the gravity dual of conformal sequestering \cite{conformalsequestering}.  In the language of the dual approximately-conformal field theory, a contribution to the soft terms of the
visible sector fields $C$
mediated by
a coupling of the form
\begin{equation}
\int d^4 \theta \,C^{\dagger} C \,{\cal{O}}_{\Delta}\,,
\end{equation}
where ${\cal{O}}_{\Delta}$ is an operator of dimension $\Delta$ in the CFT, is suppressed by a factor
\begin{equation}
M^2 \sim \left(\frac{\Lambda_{\rm IR}}{\Lambda_{\rm UV}}\right)^{\Delta-4} m_{3/2}^2 \,,
\end{equation}
with $\Lambda_{\rm IR}, \Lambda_{\rm UV}$ the infrared and ultraviolet scales, respectively, in the CFT.  In gravity language, supersymmetry breaking is mediated by perturbations to the supergravity background, and in suitable warped throat solutions -- e.g., a Klebanov-Strassler throat attached to a compact space -- these perturbations decay rapidly away from the source, sequestering the breaking of supersymmetry.

Crucially, the analysis of \cite{KMS} was performed in the no-scale limit, i.e.\
with the complex structure and dilaton stabilized by fluxes, but with the K\"ahler moduli unstabilized.
One should therefore ask whether the sequestering observed in \cite{KMS} persists upon stabilization of the K\"ahler moduli.  More specifically, the absence of superpotential cross-couplings between the visible and hidden sectors is a requirement for sequestering, as we shall review in more detail in \S\ref{sec:review}; by nonrenormalization of the superpotential it is straightforward to arrange that no such coupling arises in perturbation theory.  However, such cross-couplings 
are likely to arise at the nonperturbative level.  Nonperturbative superpotentials for the K\"ahler moduli $T$ can induce new contributions to the soft masses via interactions of the form
\begin{equation}
\Delta W =  {\cal{O}}_{vis}\, e^{-a T}\,, \label{ovis}
\end{equation} where $a$ is a constant and ${\cal O}_{vis}$ is a gauge-invariant chiral operator composed of visible sector superfields, which to cubic order in the MSSM fields can be written as
\be \label{ovisis}
 {\cal{O}}_{vis} =  \mu H_u H_d + \lambda_{ij}^{u} Q^i \q^j  H_u +\lambda_{ij}^{d}  Q^id^j  H_d+ \lambda_{ij}^{l}  L^i e^j  H_d  \,.
\ee
We have used the standard notation for the chiral superfields of the MSSM, the indices $i,j=1,2,3$ run over families, and $\mu$ and $\lambda_{ij}^{u,d,l}$ are constants that are {\it{not}} necessarily related to the tree-level $\mu$ term and Yukawa matrices, respectively (cf. \S\ref{Sec:Corrs}).  

One might be tempted to ignore nonperturbatively-small interactions, but this is not consistent in vacua for which nonperturbative effects play a critical dynamical role.  In particular, when the K\"ahler moduli are stabilized by a nonperturbative superpotential, one must ask whether this superpotential also spoils sequestering.    Answering this question is the primary goal of the present paper.

In brief, we shall find that in very simple toy models of sequestering in KKLT vacua, the nonperturbative superpotential for the K\"ahler moduli induces soft $B$ terms of order $m_{3/2}$ in a D3-brane `visible sector', spoiling sequestering, as expected from the above arguments.  In more realistic models, the gauge symmetry of the MSSM partially protects the sfermions from superpotential de-sequestering, and the flavor structure depends on the moduli stabilization scenario.
In KKLT vacua, the sfermions receive flavor-diagonal masses that are suppressed with respect to the gravitino mass. On the other hand, the Higgs and Higgsino masses receive corrections of order $m_{3/2}$.\footnote{Higgs sector
masses of this form were considered in \cite{Choi:2005uz}.}
We note that the Higgs sector is very sensitive to the details of the global compactification and is thus in no sense sequestered.
We then argue that for certain parameter regimes in the Large Volume Scenario, the corrections to the masses of the sfermions are larger, and can introduce significant flavor violation.

The outline of this paper is as follows. In \S\ref{sec:review} we critically review arguments for sequestering in supergravity and in string compactifications.
In \S\ref{sec:toy} we incorporate nonperturbative stabilization of a
K\"ahler modulus in a simple and explicit string theory  toy model.
In \S\ref{sec:realistic} we consider  nonperturbative superpotential contributions to the soft masses of a more realistic, MSSM-like,  visible sector.
We close with  conclusions in \S\ref{sec:conclusions}.  
In Appendix
\ref{smearing} we show that warped sequestering survives the relaxation of a technical assumption made in \cite{KMS}, and in Appendix \ref{conifold} we give details of the calculation of soft masses in our explicit example.


\section{Sequestering in Supergravity and String Theory}\label{sec:review}


\subsection{Sequestering from barren extra dimensions}\label{s:RS}

In \cite{RS}, Randall and Sundrum argued that locality in a higher-dimensional spacetime strongly constrains the soft terms observed in a lower-dimensional world. Their observation has three key ingredients. First of all, assuming that only the gravity multiplet propagates in the bulk, higher-dimensional locality restricts the form of the K\"ahler potential, the superpotential and the gauge kinetic function.  In \cite{RS}, this was demonstrated by considering an off-shell formulation of supergravity in which
the field $\Phi = 1 + \theta^2 F_{\Phi}$ houses some of the auxiliary degrees of freedom for the supergravity multiplet. The relevant portion of the Lagrangian is given by
\begin{align}
\frac{1}{\sqrt{-g}} \ {\cal{L}} &= \int d^4 \theta f(C^{\dagger}, e^{-V} C, \X^{\dagger},  \X) \Phi^{\dagger} \Phi + \int d^2 \theta \left( \Phi^3 W(C, \X) + \tau(C, \X) {\cal W_{\alpha}}^2 \right) \\
&-\frac{1}{6} f( \tilde{c}^*, \tilde{c}, \x^*, \x) \left({\cal{R}} + \ldots \right). \nonumber
\end{align}
Here the visible sector chiral superfields are collectively denoted by $C$ and the hidden sector
fields are denoted by $\X$, with lowest components $\tilde{c}$ and $\x$,
respectively. The visible sector vector multiplets are collectively denoted by $V$, with ${\cal W_{\alpha}}$ the gauge field strength superfields, and $\tau$ the corresponding gauge kinetic functions, while ${\cal{R}}$ is the four-dimensional Ricci scalar. The $f$ function is related to the K\"ahler potential by
\be
f = -3 M_{Pl}^2 \ e^{-K/(3 M_{Pl}^2)}  .
\ee

The {\it{assumption}} that the visible sector  communicates with the hidden sector only through the gravity multiplet implies that in the supersymmetric flat space limit, the visible and hidden sectors must decouple. Formally, in this limit ${\cal{R}}=0$ and $\Phi = 1$, so that one finds
\be
f(C^{\dagger}, e^{-V} C,\X^{\dagger},  \X)  &=& f_{\rm hid}( \X^{\dagger},  \X) + f_{\rm vis}(C^{\dagger}, e^{-V} C)\,,  \label{fseq} \\
W(C,\X ) &=& W_{\rm hid}(\X) + W_{\rm vis}(C)\,, \label{wseq}\\
\tau &=& \tau_{\rm hid}(\X) + \tau_{\rm vis}(C)\,.
\ee
We will refer to these conditions collectively as {\it separability}.
The condition (\ref{fseq}) is equivalent to the statement that the K\"ahler potential takes the special form
\be
K = - 3\Ms \ln \left( -\frac{ f_{\rm vis} + f_{\rm hid}}{3 \Ms} \right). \label{kseq}
\ee
The second observation in \cite{RS} concerns the vanishing of the tree-level soft terms. Let us define as usual the $B\mu$ term, the trilinear $A$ terms $A_{abc}$, and the soft masses $M^{2}_{a \bar{b}}$ of the MSSM as
\be\label{Lsoft}
\mathcal{L}_{soft}= M^{2}_{a \bar{b}} C^{a}\bar{C}^{\bar b}+\left( \frac{1}{2} B_{a b} C^a C^b  +\frac16 A_{abc} C^{a} C^{b} C^{c}+h.c.\right)\,,
\ee
where the visible sector fields $C^{a}$ include the Higgses, and in the MSSM
$B_{h_u h_d} = B \mu$ is the ordinary $B \mu$ term.
Then for superpotentials and K\"ahler potentials of the form\footnote{We will see in \S\ref{s:extns} that separability to all orders in the visible sector fields is an unnecessarily strong requirement, and in fact a weaker condition is sufficient to suppress the soft terms.} (\ref{fseq}) - (\ref{wseq}), one can verify that $A_{abc}=M^{2}_{a \bar{b}}=0$.\footnote{The separable structure of (\ref{fseq}) - (\ref{wseq}) does not automatically imply that the $B\mu$ term is small, but does ensure the vanishing of gravity-mediated $B\mu$ terms in the absence of supersymmetric mass terms for visible sector fields.}
Summarizing, the separable structure (\ref{fseq}) - (\ref{wseq}) combined with the absence of supersymmetric visible sector masses leads to sequestering, in that the hidden sector does not induce any soft terms in the visible sector at tree level in supergravity.

When the Planck-scale theory does not respect any flavor symmetry, acceptable flavor structure in the low-energy theory requires suppression of the gravity-mediated soft terms.  Sequestering due to extradimensional locality provides a promising mechanism for such suppression, but does not constitute a complete mediation scenario.  Instead, sequestering clears the way for  small contributions to the soft masses --- which would be overwhelmed by gravity-mediated contributions if the latter were present --- to dictate the visible sector spectrum.  The third key ingredient in \cite{RS} was the proposal that anomaly mediation could yield satisfactory soft terms in sequestered configurations.  The goal of the present paper is to investigate the possibility of sequestering in string compactifications, leaving for the future the task of constructing a mediation scenario, and so we will not have more to say about anomaly mediation.


\subsection{Corrections from moduli-mediated interactions}

A key requirement for the argument of \cite{RS} is that the extra dimensions must be barren, with only the gravity multiplet propagating in the bulk.  The proposal was that any fields not in the gravity multiplet would obtain masses at least as large as the Kaluza-Klein scale,
\be
m \gtrsim   \frac{1}{R}\,,
\ee with $R$ a typical length scale of the compactification.  This would lead to an $e^{-m R}$ Yukawa suppression of the real-space propagator of these fields, which in turn would give rise to a large suppression of any effect these fields might have on the soft terms of a visible sector separated from the hidden sector by a distance $R$.

In practice, barren extra dimensions are quite rare, both in string compactifications and in more general extradimensional model-building: compactification moduli typically induce new gravitational-strength interactions that mediate supersymmetry breaking and hence spoil sequestering.  The authors of \cite{Anisimov:2002az} examined a variety of string theory models with calculable K\"ahler potentials, and found that the special form (\ref{kseq}) does not seem to be generic in M-theory or string theory with branes, despite the manifest extradimensional locality of such models.  The unwanted couplings arise from the exchange of bulk supergravity fields, particularly moduli.

Along the same lines, the authors of \cite{Kachru:2006em} gave a general argument showing that the assumption of barren extra dimensions does not hold in string compactifications,  even after stabilization of the  moduli.  In any compactification for which moduli stabilization can be described in the four-dimensional effective theory, the moduli masses will be no larger than the cutoff scale of the four-dimensional effective theory, and in particular will not exceed the Kaluza-Klein scale, so that $mR \ll 1$.  Thus, the effects transmitted by massive, stabilized moduli cannot be neglected in general.

The importance of moduli stabilization for sequestering was first emphasized by Luty and Sundrum in \cite{LS}.  They considered five-dimensional supergravity compactified on $S^{1}/{\mathbb Z}_{2}$ and asked whether supersymmetry breaking on the hidden orbifold boundary gave rise to sequestered supersymmetry breaking for matter fields on the visible brane.  Prior to stabilization of the radion controlling the interval size, the K\"ahler potential took a sequestered form.  To stabilize the radion, they invoked gaugino condensation in a bulk gauge group and in a boundary gauge group, yielding a superpotential
\begin{equation} \label{LSeq}
W = c + b \,e^{-a T} \,,
\end{equation}
for constants $a, b, c$, and with $T$ the radion.  Luty and Sundrum then showed that with this superpotential, sequestering survives the stabilization of the radion.

In string theory, the situation is somewhat more complicated, for several reasons.  To assist the reader in navigating the remainder, we briefly sketch these complications.  First, even before stabilization of the moduli, the K\"ahler potential for an unwarped string compactification does {\it{not}} generically take the sequestered form (\ref{kseq}),
as we have just reviewed.  However, strong warping ameliorates some of the moduli-mediated interactions, as we shall explain in \S\ref{KMSsection}. Moreover, a criterion weaker than \eqref{fseq} for the suppression of K\"ahler potential couplings appears well-motivated in certain unwarped examples, and may allow effective sequestering even in compactifications violating \eqref{fseq} (\S\ref{s:extns}).  Most importantly, nonperturbative stabilization of the K\"ahler moduli introduces new interactions that violate \eqref{wseq} and spoil sequestering, even in the presence of warping, as we will explain in \S\ref{desequester}.


\subsection{Sequestering in warped compactifications \label{KMSsection}}

Although moduli-mediated interactions render sequestering non-generic in unwarped string compactifications \cite{Anisimov:2002az}, some of the problematic effects are suppressed by strong warping \cite{KMS}.  Suppose that the
supersymmetry-breaking sector is localized at the bottom of a warped throat.
From the ten-dimensional perspective, the mediation of supersymmetry breaking to a visible sector some distance up (or even outside) the throat will proceed through perturbations of the supergravity fields sourced in the infrared, i.e.\ from the bottom of the throat. Taking the throat to be a warped Calabi-Yau cone with Sasaki-Einstein base $X_{5}$, the bulk fields $\varphi$ can be expanded in eigenmodes on $X_{5}$, so that schematically one has
\be
\varphi = \sum_{\alpha} c_{\alpha} r^{- \Delta_{\alpha}} Y_{\alpha} (\Psi).
\ee
Here $\alpha$ indexes the quantum numbers under the isometries of $X_{5}$, $c_{\alpha}$ are constants, $r$ is the radial coordinate, $\Psi$ denotes the angular coordinates on $X_{5}$, $Y_{\alpha}$ is an angular harmonic on
$X_{5}$, and $\Delta_{\alpha}$ is the dimension of the operator that is dual, via AdS/CFT, to the corresponding supergravity mode. Couplings between a supersymmetry-breaking sector located at the tip of the throat and a visible sector located at the top of the throat are suppressed by powers of the hierarchy of scales in the throat. When all operators inducing cross-couplings have $\Delta > 4$,
the gravity-mediated soft terms are highly suppressed, and the system experiences {\it{warped sequestering}}, which is the gravity dual of conformal sequestering \cite{conformalsequestering}.

\subsubsection{Warped sequestering in the Klebanov-Strassler throat \label{sec:WarpedSeq}}
Let us briefly review and clarify the results of \cite{KMS},
which analyzed the mediation of supersymmetry breaking from an anti-D3-brane to a D3-brane by normalizable profiles of the supergravity fields in a Klebanov-Strassler throat region of a
no-scale compactification.   The leading effects arose from the lightest Kaluza-Klein modes on $T^{1,1}$ --- and correspondingly the lowest-dimension operators in the Klebanov-Witten theory \cite{Klebanov} --- with two properties: the mode must be sourced by an anti-D3-brane, and it must induce supersymmetry-breaking
masses for the D3-brane fields.  A particular supergravity field, denoted by $\Phi_-$ in the conventions of \cite{BDKMS}
(their equation (2.5)), controls D3-brane scalar masses, and so the task was to determine the lowest modes in the spectrum of $\Phi_-$ excitations that are sourced by an anti-D3-brane.\footnote{We implicitly assume that the anti-D3-brane configuration corresponds to a supersymmetry-breaking state of the cascading gauge theory, as in \cite{KPV}, but it would be valuable to confirm or exclude this along the lines of \cite{BGH}.}


In \cite{KMS} (to which we refer for explanation of our notation) it was argued, following \cite{Ceresole}, that the lowest-dimension operator dual to a normalizable mode of $\Phi_-$ is a non-chiral operator ${\cal O}_{\sqrt{28}}$ with quantum numbers
${\bf (3,3,1)}$ under the $SU(2) \times SU(2) \times U(1)$ global symmetry, and with dimension $\Delta=\sqrt{28} \approx 5.29$.  However, upon comparing to the result of \cite{KKLMMT} for the Coulomb potential between a D3-brane and an anti-D3-brane in a warped throat, the authors of \cite{KMS} found that this mode is apparently not sourced by an anti-D3-brane, and the leading mode of $\Phi_-$ in the actual solution is dual to
the non-chiral singlet ${\cal{O}}_{8} = \int d^4 \theta \, {\rm tr}\left({\cal W_{\alpha}} {\cal W^{\alpha}}\overline {\cal W}_{\dot \beta}\overline {\cal W}^{\dot \beta}\right)$ with dimension $\Delta = 8$.  This result, which was subsequently confirmed in \cite{DKM,BGGH},
might lead one to expect that there exists a different supersymmetry-breaking state (potentially preserving different global symmetries) in which the apparently more relevant operator ${\cal O}_{\sqrt{28}}$ obtains an expectation value and leads to less-sequestered soft terms.

This expectation would be erroneous:
although the operator ${\cal O}_{\sqrt{28}}$ is indeed present in the Klebanov-Witten theory, the vev of ${\cal O}_{\sqrt{28}}$ is {\it{not}} dual to a normalizable perturbation of $\Phi_-$, and moreover  the lowest-dimension operator whose expectation value is dual to a normalizable perturbation of $\Phi_-$ is precisely the operator ${\cal{O}}_{8}$ induced by an anti-D3-brane. Therefore, the anti-D3-brane activates the most relevant $\Phi_-$ perturbation available in the theory, and moreover this mode is a singlet, as one would expect at leading order in a
multipole expansion.

To correct the assignments of operators to supergravity modes, we refer to the discussion in \S3.3 of \cite{BDKMS}.  There it was observed that a mode of $\Phi_-$ dual to the vev of an operator with dimension $\Delta$ has a radial profile $\delta \Phi_- \propto r^{4-\Delta}$, which differs by a factor $r^{4}$ from the result for a scalar field in $AdS_{5}$ with the standard normalization.  As a result, a given operator in the Klebanov-Witten theory whose expectation value is dual to a normalizable mode of $\Phi_-$ will have its source dual to a non-normalizable mode of an independent supergravity field \cite{BDKMS}, which we denote $\Phi_+$.   In particular, as explained in \cite{BDKMS}, the simplest operators dual to normalizable $\Phi_-$ profiles are of the form ${\rm Tr}\left({\cal W_{\alpha}} {\cal W^{\alpha}}\overline {\cal W}_{\dot \beta}\overline {\cal W}^{\dot \beta}(AB)^{k}\right)$, with $k$ a non-negative integer.  The lowest operator in this tower is the $\Delta = 8$ singlet described above.

What of the operator ${\cal O}_{\sqrt{28}}$?  Its expectation value is dual to a normalizable mode of $\Phi_+$ (while a source for this operator is dual to a non-normalizable mode of $\Phi_-$). Therefore, a vev of this operator does not induce soft terms for a D3-brane visible sector in the ultraviolet.

However, there is a mode with much lower dimension that could potentially induce soft terms at the nonlinear level: this is a mode of flux dual to the chiral operator ${\cal{O}}_{5/2}= \int d^2 \theta \, {\rm tr}(A^{i}B^{j})$ with quantum numbers ${\bf (\frac{1}{2},\frac{1}{2},-1)}$ and dimension $\Delta=5/2$ (cf.\ \cite{fluxpaper}).  In Appendix \ref{smearing} we demonstrate that this mode does not alter the conclusions of \cite{KMS}.

In conclusion, for anti-D3-brane supersymmetry breaking in the Klebanov-Strassler solution, the lowest-dimension operator mediating soft terms to a D3-brane `visible sector' has $\Delta = 8$, so that the sequestering is very strong.

The considerations described above are strictly applicable only to a noncompact Klebanov-Strassler throat.  For a finite throat region attached to a compact space, there is  at least one new light degree of freedom,
the K\"ahler modulus $T$ controlling the overall volume.  Because $T$ is not part of the CFT, it is natural to expect that $T$ will mediate soft terms that are {\it{not}} suppressed by the hierarchy of energy scales in the throat.
In \S\ref{desequester} we will observe that nonperturbative superpotentials for the K\"ahler moduli indeed generically spoil warped sequestering.  However, we first describe a weaker criterion for sequestering which is applicable in certain cases with a well-localized visible sector.


\subsection{Sort-of sequestering}\label{s:extns}
We have reviewed how extradimensional locality and the assumption of barren extra dimensions imply the separability
of $W$ and $f$, as in \eqref{fseq} and \eqref{wseq}, and how \eqref{fseq} and \eqref{wseq} in turn imply sequestering of supersymmetry breaking.
Because light moduli in string compactifications violate \eqref{fseq}, it is reasonable to ask whether a weaker  assumption might suffice to ensure suppression of the soft masses in comparison to $m_{3/2}$.

To identify this weaker condition, we expand a general K\"ahler potential and superpotential in powers of the visible sector fields $C^{a}$ as
\be
K & = & \widehat{K}(\X, \bar{\X}) + \tilde{K}_{a \bar{b}} (\X, \bar{\X}) C^{a}
\bar{C}^{\bar{b}}+ \left[ Z(\X, \bar{\X}) H_u H_d + h.c. \right] \ldots \\
W & = & \widehat{W}(\X) + \mu(\X) H_u H_d + \frac{1}{6} Y_{a b c}(\X) C^{a} C^{b} C^{c} + \ldots  \label{wis}
\ee
and then require the separability of the $f$ function \textit{only at leading order}\footnote{In terms of the notation of \S\ref{s:RS} this means that $f$ can have a part $f_{\rm mix}$ involving both the hidden sector fields and the visible sector fields as long as it satisfies
$\partial_a f_{\rm mix} (X, \bar{X}, C, \bar{C}) = \partial_a \bar{\partial}_{\bar b} f_{\rm mix} (X, \bar{X}, C, \bar{C}) = 0$.} in the visible sector fields \cite{Blumenhagen:2009gk}. This condition, which is related to the ``extended no-scale structure'' in \cite{Blumenhagen:2009gk}, reads
\be\label{extns}
\tilde{K}_{a \bar{b}}=e^{\widehat{K}/3 M_{Pl}^2} \kappa_{a\bar b}\,,
\ee
with $F^{m}\partial_{m} \kappa_{a\bar b}=0$, i.e.~$\kappa_{a\bar b}$ does not depend on the moduli that get non-vanishing F-term vevs. In particular this means that one can always rotate and rescale\footnote{For a generic moduli dependence, it is possible to diagonalize $\tilde{K}_{a \bar{b}}(\X, \bar \X)$ only at a single point in the moduli space. As computing
the
soft terms \eqref{Bsoft}-\eqref{Aterm} requires differentiating with respect to the moduli, this is not sufficient.} the $C^{a}$ such that $\kappa_{a\bar b}\rightarrow \delta_{a\bar b}$. Using \eqref{extns} and the standard supergravity formulae \cite{KL,Brignole:1997dp,Soni:1983rm} it is easy to verify that a series of cancellations leads to the following result for the soft terms of the MSSM fields (not yet canonically normalized, which we emphasize  with the hats):
\be
\hat M^{2}_{a \bar{b}}&=&\frac23 \frac{V_{0}}{M_{Pl}^{2}} \tilde K_{a\bar b} \simeq 0\,,\label{Mvic}\\
\hat A_{abc}&=& e^{\widehat K/2\Ms} F^m \partial_m Y_{a b c}\,,\label{Avic}\\
B\hat \mu &=& e^{\widehat K/2 M_{Pl}^{2}} \mu \left[F^{m} \left(\partial_{m}\log \mu -\frac{\widehat W_{m}}{3 \widehat W}\right)\right]+\mu \frac{V_{0}}{3\widehat W}+\mathcal{O}(Z)\,,
\ee
where $V_{0}$ is the vacuum energy at the minimum of the F-term potential, which we are assuming is negligibly small, and $\mathcal{O}(Z)$ stands for terms proportional to $Z$ and its derivatives, which we omit for simplicity. It is clear from \eqref{Avic} that in order to ensure the absence of gravity/moduli mediated $A$ terms\footnote{For $B\mu$ even the separability of both $f$ and $W$ is not sufficient to ensure sequestering. Instead one needs to make further assumptions, e.g.\ the absence of supersymmetric visible sector masses, i.e.~$\mu=Z=0$.} one needs to  assume the separability \eqref{wseq} of the superpotential, such that $F^m \partial_m Y_{a b c}=0$.

As for the all-orders separability in \eqref{fseq} and \eqref{wseq}, the leading-order separability condition \eqref{extns} is also not generically satisfied in string compactifications. Nevertheless, there are arguments that \eqref{extns} might be valid, at least approximately.
In fact, in \cite{Conlon:2006tj} it was argued
that the combination of locality and holomorphicity enforces a special form of the metric on the visible sector moduli space. Their argument, which we will review in \S \ref{sec:LVS}, suggests that $K_{a\bar b}\sim e^{K/3\Ms} \kappa_{a\bar b}$ with $\kappa_{a\bar b}$ independent of the K\"ahler moduli. When all other moduli have only small or vanishing F-terms then interesting suppressions of the soft terms as in \eqref{Mvic} and \eqref{Avic} might arise.


\subsection{Superpotential de-sequestering} \label{desequester}

Even when the separability \eqref{fseq} of $f$ can be justified in a scenario with barren extra dimensions, or when the weaker criterion \eqref{extns} follows from locality and holomorphicity, the separability \eqref{wseq} of $W$ is necessary to prevent flavor violation in the soft trilinear $A$ terms.

We now make a critical observation: the separability of $W$, \eqref{wseq}, is generically violated in string compactifications stabilized by nonperturbative effects, and the resulting soft terms therefore require careful study. This is one of the main goals of the present paper.

The best-understood scenarios for complete moduli stabilization in type IIB string theory \cite{Kachru:2003aw, BBCQ, Braun} incorporate nonperturbative contributions to the superpotential,
e.g. from gaugino condensation on a stack of D7-branes wrapping a four-cycle, to lift
the K\"ahler moduli.
Consider a visible sector residing on (possibly fractional) D-branes in a compactification of type IIB string theory.
Suppose that a K\"ahler modulus $T$ describing the volume of some four-cycle $\Sigma$ is stabilized by gaugino condensation in a super Yang-Mills sector on $N_{c}$ D7-branes wrapping $\Sigma$.
Even if $\Sigma$ is distant from the D-branes constituting the visible sector, strings stretching between the hidden and visible D-branes carry charges under both sectors.
Integrating out these strings will generically induce couplings between the sectors.
This computation has been performed explicitly in toroidal orientifolds (i.e.\ without warping), with the result that D-branes distant from the hidden sector give unsuppressed threshold corrections to the nonperturbative superpotential \cite{Berg:2004ek}.
An important question is whether these stretched strings can be massive enough to decouple if the hidden and visible sectors are well-separated along a warped direction.  A precisely analogous question arises in D3-brane inflation, in which significant contributions to the inflaton potential arise from strings stretched between the inflationary D3-brane and the D7-branes whose strong gauge dynamics stabilizes the K\"ahler moduli.  Explicit computation of the nonperturbative superpotential has revealed that the induced cross-couplings are {\it{not}} negligible, even in strongly warped backgrounds \cite{BDKMMM}.  The physical explanation for this was provided in \cite{Baumann:2006cd}, which showed that  in any warped throat with Sasaki-Einstein base, the mass of a string stretching up the throat is small compared to the four-dimensional Planck mass.

We therefore expect that nonperturbative stabilization of the K\"ahler moduli can induce new contributions to the soft masses via superpotential interactions\footnote{We remark that perturbative nonrenormalization cannot forbid nonperturbative couplings of this form.  Moreover, as long as $T$ is not charged under the symmetries of the Standard Model, these symmetries also cannot forbid the couplings (\ref{ovis}).} as in equation \eqref{ovis},
$\Delta W =  {\cal{O}}_{vis} \, e^{-a T}$.

Nonperturbative stabilization of the radion was also considered in \cite{LS}, and was found to be compatible with sequestering.  The critical difference between \cite{LS} and the present work is that we allow couplings to the visible sector in the nonperturbative superpotential for the K\"ahler moduli, so that $b$ in (\ref{LSeq}) would be a gauge-invariant combination of visible sector fields, cf.\ (\ref{ovisis}), rather than a pure constant.  Clearly,  this  dramatically changes the physical outcome.  Let us stress that the superpotential (\ref{LSeq}) of
\cite{LS} does follow upon assuming the absence of couplings between the hidden and visible sectors.
However, {\it{in string compactifications for which the hidden and visible sectors are composed of D-branes}}, one invariably has a spectrum of massive strings stretching between these D-branes, and integrating out these strings induces cross-couplings of the form (\ref{ovis}).



\section{The Effects of Moduli Stabilization: A Toy Model}\label{sec:toy}

We have argued above that in a compactification whose K\"ahler moduli are stabilized by a nonperturbative superpotential, superpotential cross-couplings (\ref{ovis}) between the visible sector and the K\"ahler moduli induce soft supersymmetry breaking in the visible sector.  To assess the form of the resulting soft terms, we turn to  a string theory toy model with stabilized moduli in which the resulting soft terms can be computed explicitly.

\subsection{Supersymmetric vacuum for a D3-brane} \label{SectKKLT}

As in \cite{KMS}, we consider a D3-brane in a Klebanov-Strassler throat \cite{Klebanov:2000hb}.
The D3-brane will serve as a proxy for the visible sector, not because its low-energy effective theory gives a good approximation to the phenomenological features of the standard model, but because it is a simple but nontrivial case where one can test the warped sequestering proposal.  
First, in \S\ref{sec:KKLTBkg}, we recall the essentials of KKLT moduli stabilization \cite{Kachru:2003aw} and then, in \S\ref{explicittoy}, we obtain a supersymmetric vacuum\footnote{The solution that we will investigate in detail was first noticed in \cite{DeWolfe:2007hd}, but we mention in \S\ref{AppSol} how this extends to a broader class of solutions.} for a D3-brane in the conifold.  The soft terms induced by supersymmetry breaking are then obtained in \S\ref{softsection} and evaluated in \S\ref{BoundBSect}. In \S\ref{Lessons} we extract
a few lessons from the toy model.

\subsubsection{The KKLT scenario} \label{sec:KKLTBkg}

In warped compactifications of type IIB string theory on conformally Calabi-Yau three-folds with flux and O-planes, the four-dimensional effective theory contains complex structure moduli, K\"ahler moduli, and the axio-dilaton,
as well as open string moduli
due to branes. For simplicity, we will consider the case of a single K\"ahler modulus $T$, and a single spacetime-filling D3-brane whose position
is parameterized by three complex scalars $\phi_{i}$, $i=2,3,4$.
Three-form flux generates a classical superpotential for the axio-dilaton and complex structure moduli,
\be
W_0 = \int G_3 \wedge \Omega,
\ee  providing  these moduli with reasonably large masses. After integrating out the massive moduli\footnote{The K\"ahler potential including the complex structure  moduli and dilaton is not of the form (\ref{kseq}), and thus one might be tempted to conclude that these models do not sequester. However, in the case that the complex structure moduli and axio-dilaton do not obtain large F-terms after supersymmetry breaking \cite{Choi:2008hn}, these moduli are not part of the supersymmetry-breaking hidden sector, and the condition (\ref{kseq}) need not apply to these moduli for sequestering to work.}
and introducing a nonperturbative superpotential from gaugino condensation on $n$ D7-branes, the $\N = 1$ effective theory for the K\"ahler modulus and D3-brane matter fields takes the form:
\be
K &=& -3 \Ms \ln \Bigl(T +\bar{T} - \gamma k(\phi,\bar{\phi})\Bigr) \equiv -3 \ln U, \label{KKKLT} \\
W &=& W_0 + W_{\rm np} = W_0 + {\cal{A}}(\phi) e^{-a T}\,, \label{KKLTw}
\label{KKLT}
\ee where $k(\phi,\bar{\phi})$ is the  K\"ahler metric on the Calabi-Yau manifold, $\gamma = \frac{1}{3 \Ms}$, $U = T +\bar{T} - \gamma k(\phi,\bar{\phi}) = {\cal V}^{2/3}$,
and $a = \frac{2 \pi}{n}$.

Before delving into the particulars of explicit models, we readily observe that the nonperturbative effect responsible for the stabilization of the D3-brane and the K\"ahler modulus involves a direct cross-coupling between these two sectors, as in equation (\ref{KKLTw}). If the D3-brane is regarded as a toy visible sector, a large F-term for
$T$ would indeed violate the separability condition (\ref{wseq}) for sequestering.  We now turn to quantifying this contribution to the D3-brane soft mass.

\subsubsection{Supersymmetric D3-branes in the conifold }\label{explicittoy}

To incorporate moduli stabilization explicitly, we follow \cite{BDKMMM,BDKM} and embed a stack of $n$ D7-branes in the throat region, along a divisor $z_{2}=\mu$ \cite{Kuperstein}. Here $\mu$ is a complex constant that encodes the D7-brane location, and we are using the standard coordinates in which the deformed conifold is defined by the locus
\be \label{Defcon}
\sum_{A = 1}^4 z_A^2  = \epsilon^2
\ee in
${\mathbb C}^4$.
Gaugino condensation on the D7-branes then yields the nonperturbative superpotential \cite{BDKMMM, Berg:2004sj}
\begin{equation} \label{d3w}
W_{\rm np} = {\cal{A}}_0 \Bigl(\frac{z_2 - \mu} {\mu}\Bigr)^{1/n} e^{- a T}\,,
\end{equation} where ${\cal{A}}_{0}$ is a constant with dimensions of
 $(\rm{mass})^{3}$.
Before incorporating the effects of supersymmetry breaking, we  obtain a supersymmetric AdS solution by solving the F-flatness equations
for the ansatz (\ref{KKKLT}), (\ref{KKLTw}):
\begin{align}
D_{T} W &= 0 \label{susyrho}\,, \\ 
D_i W &= 0  \label{susyD3susyD3} \; .
\end{align}
Using (\ref{susyrho}), the F-flatness conditions (\ref{susyD3susyD3}) for the three independent D3-brane coordinates $z_{i}$, $i = 2, 3,4$, can be written
as
\be
\partial_i \left[\ln\left(\frac{z_2 - \mu} {\mu}\right)\right] - a n \gamma k_i = 0 \,,\label{D3Susy}
\ee which, we observe, does not depend on $T$.  This makes it convenient to first find the position of the D3-brane, and then feed this information into (\ref{susyrho}).
Each solution has a two real-dimensional moduli space, consistent with the residual $SO(2)$ symmetry of the solution.

We now expand in $|\frac{z_2}{\mu}| \equiv \frac{1}{B}$, keeping only the leading term, to obtain a solution in which the D3-brane is at much smaller radial position than the D7-brane, but is still far above the tip.
As we will demonstrate in \S\ref{BoundBSect}, this leads to the hierarchy of scales
\be
0 \approx \left|\frac{\epsilon}{\mu}\right| \ll \left|\frac{z_2}{\mu}\right| \ll 1,
\ee and we can consistently set $\epsilon = 0$. We choose to study a D3-brane localized at a specific point in this moduli space, which to this order is defined by $z_{3}=z_{4} = 0$ and
\begin{align}
z_2 = \frac{|\mu|}{B}  =  \frac{1}{ 4 \left( \frac{4 \pi \gamma}{3}   \right)^3 |\mu|^3} \,,
\end{align}
where we have taken $z_2$ real.
In this class of solutions we have $r^3  = 2 |z_2|^2$, and $z_1 = \pm i z_2$ is purely imaginary. In our discussion of the masses we will consider the upper sign. Once the location of the D3-brane is found, (\ref{susyrho}) becomes a single-variable transcendental equation, which can be solved numerically. We now turn to the effects of supersymmetry breaking on this vacuum.

\subsection{Soft masses for the D3-brane} \label{softsection}

We will break supersymmetry, as in KKLT, by adding an anti-D3-brane at the tip of the throat, which contributes the `uplift' potential
\be
V_{\rm up} = \frac{ D}{(T +\bar{T} - \gamma k(z,\bar{z}))^2}\,, \label{vup}
\ee
where
the constant $D$ is determined by requiring the approximate cancellation of the cosmological constant.  The results of \cite{KMS} ensure that higher-order terms in the expansion of the brane-antibrane potential are negligible.

The supersymmetry-breaking contribution (\ref{vup}) to the potential for $T$ and $\phi_{i}$ will in general induce shifts in the vevs.  Let us use $Z^{M}$ to denote the vevs of $T,\bar T,z_{i},\bar z_{i}$, with
$Z_0^M$ denoting the vevs in the supersymmetric solution, and $\Zs^M \equiv Z_{0}^{M} + \delta Z^{M}$ denoting the vevs in the supersymmetry-breaking solution.  We will be interested in configurations for which $\delta Z^{M} \ll Z_{0}^{M}$, i.e. the shifts induced by the uplifting are small.  In this approximation,
we can obtain the shifts $\delta Z^{M}$
by solving 
\be
0 = \partial_{M} (V_F + V_{up}) \vline_{ \ \Zs} \approx \partial_M V_{up} \axs + \delta Z^N \left[ \nabla_N \nabla_M ( V_F + V_{up} ) \right] \axs\,.  \label{taylor}
\ee
This amounts to inverting the total mass matrix evaluated at the supersymmetric point,  $\delta Z^M = -   \left( \partial_N \partial_M ( V_F + V_{up} ) \right)^{-1}  \axs \partial_N V_{up} \axs$.
One finds (see
\S\ref{AppMass} for details, including justification for dropping
covariant derivatives) that the volume modulus shifts to a slightly larger value.  However, the D3-brane remains at the point where it was supersymmetrically stabilized, to leading order in $1/B$ and to quadratic order in $\frac{1}{aU}$, despite the fact that $V_{\rm up}$ depends on the D3-brane coordinates.

We can now obtain the mass matrix for the supersymmetry-breaking vacuum
in a Taylor expansion around the supersymmetric point: at leading order we find
\begin{align}
\nabla_{M}\nabla_{N} (V_{\rm tot}) \vline_{\,Z_{\star} } 
&= \partial^2_{MN} \left( V_{\rm F} + V_{\rm up} \right) \vline_{\,Z_{\star} }  \\
&= \partial^2_{MN} \left( V_{\rm F} + V_{\rm up} \right) \axs + \delta Z^P\,\nabla_P \partial^2_{MN} \left(  V_{\rm F} + V_{\rm up} \right) \axs  \label{expansion}.
\end{align}
In fact, the contribution proportional to $\delta Z^{T}$ turns out to be negligible in this example: explicit evaluation of
\be \label{small}
(\Delta M^2_{\rm tot})_{MN} = \delta Z^P\,\nabla_P \partial^2_{MN} \left(  V_{\rm F} + V_{\rm up} \right) \axs
\ee
shows that these contributions are suppressed
by
at least one power of
$1/(aU)$ or $1/B$ with respect to the zeroth-order contribution in the expansion of the mass matrix.
Thus, to lowest order in  $1/B, 1/(aU)$, the mass matrix is given by the full mass matrix evaluated at the supersymmetric point,
\be
\partial^2_{MN} \left( V_{\rm tot} \right) \axst \approx \partial^2_{MN} \left( V_{\rm tot} \right) \axs \,.
\ee
We
refer the interested reader to
\S\ref{AppMass} for the full details.

Now, the scalar masses at the supersymmetric point ($D_A W = 0$) 
are easily
obtained by differentiating the F-term potential (\ref{pot}) twice,
\begin{align}
\partial_a \partial_{\bb} V_F  \axs &= e^{K/ \Ms} \left(  \KAB \partial_a (D_A W) \partial_{\bb} (\bar{D}_{\bar{B}} \overline{W})   - 2 \frac{|W|^2}{M^4_{Pl}} K_{a \bb} \right) \,, \label{SusyMass}  \\
\partial_a \partial_{b} V_F  \axs &= - \frac{e^{K/\Ms} \overline{W}}{ \Ms} \partial_{ a } D_{ b }W. \label{SusyHolMass}
\end{align}
The holomorphic masses in (\ref{SusyHolMass}) would not be present in unbroken rigid supersymmetry,
where  holomorphic masses only appear in the form of
soft $B$ terms. In local supersymmetry, however, these terms can be nonzero without breaking supersymmetry, as discussed in e.g. \cite{Wess:1992cp}. The ordinary masses in equation (\ref{SusyMass}) are better understood when considered in conjunction with the fermion masses evaluated at the supersymmetric point,
\begin{align}
m_{ab} \Bigl|_{Z_{0}} &=  e^{\frac{K}{2 \Ms} }   \partial_a D_b W\,, \\
\mgrs \axs &= e^{K/\Ms} \frac{|W|^2}{\M^4},
\end{align}
so, expressed in terms of fermion masses evaluated at the supersymmetric point, we have
\be
\partial_a \partial_{\bb} V_F \axs &= K^{c \bar{d}} m_{ac} m_{\bb \bar{d}} - 2 \mgrs K_{a \bb}\,.
\ee
The uplift  potential (\ref{vup}) induces supersymmetry-breaking masses of the form
\begin{align}
\partial^2_{M N} V_{up} = \frac{2 V_{\rm up}}{3 \Ms}\left( K_{M N} + \frac{2}{3 \Ms}K_M K_N \right)\,, \label{mup}
\end{align}
where $M, N$ can be either holomorphic  or antiholomorphic.
Using that $V_{\rm tot} \axs = (V_F + V_{\rm up})\axs \simeq 0$, we can write
\begin{align}
\partial^2_{a \bb} V_{\rm tot} \axs &=  K^{c \bar{d}} m_{ac} m_{\bb \bar{d}} + \frac{4}{3} \frac{\mgrs}{\Ms} K_a K_{\bb}\,, \\
\partial^2_{a b} V_{\rm tot} \axs &= \mgrs \left[K_{ab} + \frac{4}{3 \Ms} K_a K_b - \frac{\Ms W_{ab}}{W} \right]  \; .
\end{align}
The fermion mass-squared now appears in the expression for the
nonholomorphic scalar mass-squared, which
will be useful when we consider mass splittings.

\subsection{Evaluated soft masses \label{BoundBSect}}
In 
\S\ref{AppMass}, we express our results for the mass matrix in terms of $V_F, aU$, and $B$. While the first two of these quantities are determined completely  once ${\cal A}_0, W_0$ and $a = \frac{2 \pi}{n}$ are specified, the latter is a parameter of the solution describing the position of the D7-brane in the throat.
In this section,
we explicitly evaluate the mass matrices for a particular set of values for $W_0, {\cal A}_0, n$ and $B$.
We caution the reader not to attach undue weight to the precise numbers presented here, which
serve only to allow comparison of various contributions to the soft masses.

We will consider the case $|{\cal A}_0| = \M^3$, $|W_0| = 10^{-13} \M^3$, and $a = \frac{2 \pi}{32}$, for which $aU \approx 66$.
%
%
The gravitino mass is then given, to lowest order in $1/(aU)$ and in $1/B$, by
\be
m^2_{3/2}\Bigr|_{Z_{*}}  \simeq \left|\frac{W_0}{\Ms}\right|^2 e^{K/\Ms}\Bigr|_{Z_{0}} \approx  1.3 \cdot  10^{-31} \Ms \approx ( 867\,  {\rm GeV})^2\,.
\ee
We now specialize to the case\footnote{See 
\S\ref{AppBoundB} for consideration of microphysical upper bounds on $B$.} $B = 400$, still using $aU \approx 66$ as above, so that working to lowest order in an expansion in these two quantities provides sufficient accuracy.

The eigenvalues of the full scalar mass-squared matrix at the previously supersymmetric point are $ (aU)^2 \mgrs \approx 4400\  \mgrs $ with multiplicity 
two, corresponding to the real and imaginary parts of the K\"ahler modulus. There are two flat directions, which in this case are in the imaginary $z_3, z_4$ directions. The corresponding real directions both have masses $2 \mgrs$. The last eigenvectors are almost aligned with the real and imaginary parts of the $z_2$ direction. While the imaginary direction has a mass of $\frac{3}{4} \mgrs$, the real part is tachyonic with a mass-squared of $-\frac{1}{4} \mgrs$.
Several tachyons with mass-squared above the Breitenlohner-Freedman bound were present in the AdS vacuum, but most of these obtained nonnegative masses after uplifting.
It is possible that this last tachyon is cured in regions of parameter space that are not accessible to our perturbative expansion in $1/(aU)$ and $1/B$, but as our primary goal is to understand the transmission of soft masses rather than to construct a fully realistic model, we will not consider this point further.

The masses of these real fields will be interpreted in part as supersymmetric masses in Minkowski space, and in part as soft masses. We identify the soft part of the ordinary, non-holomorphic, masses as the mass splittings between bosons and fermions in Minkowski space, while any holomorphic mass term is regarded as a soft $B$ term mass. Evaluating the normalized soft masses, we obtain
\begin{align}
 M^{2}_{a \bar{b}}\Bigr|_{Z_{*}} &= \partial^2_{a \bb}( V_F + V_{\rm up})\Bigr|_{Z_{0}} - m_{ab}K^{b \bar{c}} m_{\bar{c} \bb}\Bigr|_{Z_{0}}
 = \frac{4 \mgrs}{3 \Ms}K_a K_{\bb}\Bigl|_{Z_{0}}   \label{MassSplit}  \\
 &= 4 \mgrs
 \left(
\delta_{ a\bb}^{\rho \bar{\rho}} - \frac{\sqrt{3}}{2 (\pi U B)^{1/2} } (\delta_{ a\bb}^{\rho \bar{2}} + \delta_{ a\bb}^{2 \bar{\rho}}) + \frac{3}{4 \pi U B}  \delta_{ a\bb}^{2 \bar{2}}
 \right)  \\
&= 4 \mgrs
 \left(
  \begin{array}{c c c c}
 1 & - 1.3 \cdot 10^{-3}  \\
- 1.3 \cdot 10^{-3}   & 1.8 \cdot 10^{-6} \\
  & & 0 \\
  & & & 0
 \end{array}
 \right)\,. 
\end{align}
The normalized soft $B$ terms are evaluated to
\begin{align}
B_{ab} &\equiv \partial_{ab} V_{\rm tot}\Bigr|_{Z_{*}}
= \mgrs \left[K_{ab} + \frac{4}{3 \Ms} K_a K_b - \frac{\Ms W_{ab}}{W} \right]  \label{B-term} \\
&= \mgrs
\left(
aU \delta_{ a b}^{\rho \rho} -  \frac{\sqrt{3  }a }{2}\left(\frac{U }{\pi B}\right)^{1/2} \delta_{ \{ a b \}}^{ \rho 2 }  -\frac{1}{2}   \delta_{ a b}^{2 2} + \delta_{ a b}^{4 4} + \delta_{ a b}^{3 3}
\right)
 \\
&= \mgrs
\left(
\begin{array}{c c c c}
66 & -  8.8 \cdot 10^{-2} \\
- 8.8 \cdot 10^{-2} & -\frac{1}{2} \\
& & 1 \\
& & & 1 \\
\end{array}
\right).
\end{align}

For completeness, we also list the F-term vevs at the non-supersymmetric solution to the lowest non-vanishing order in the perturbative expansion,
\begin{align}
F_a \axst &= D_a W \axst = D_a W \axs + \delta Z^M \nabla_M D_a W \axs  \\
&= \delta X^{T} \left( \partial_{T} D_a W + \frac{K_{\bar{T} a}}{\Ms} W \right)\axs. \label{FT}
\end{align}
For the canonically normalized volume modulus, we obtain $|F_{T} |^{1/2}\approx    \left( \frac{\sqrt{3 U}}{a} m_{3/2} \M \right)^{1/2} \approx  5.8 \cdot 10^{11}\, {\rm GeV}$. In the visible sector, the canonically normalized field corresponding to $z_2$ gets an F-term with $|F_2|^{1/2}  \approx \left(m_{3/2} \M \frac{3}{2a} \frac{1}{\sqrt{\pi B} }   \right)^{1/2} \approx 2.1 \cdot 10^{10}\, {\rm GeV}$.


\subsection{Lessons from the toy model \label{Lessons}}

Our explicit computation demonstrates that when including the effects of moduli stabilization, even a `visible sector' separated from the supersymmetry breaking by a highly warped region can acquire substantial soft masses, primarily through $B$ terms induced by the nonperturbative superpotential.

We now briefly show how
the supersymmetry-breaking mass  splittings in the visible sector can be interpreted as stemming from an array of different effects.  The divisions presented here are somewhat artificial, but can be helpful in tracing the origin of the soft masses; we will give a unified treatment in \S\ref{sec:realistic}.
First, supersymmetric Bose-Fermi mass splittings  in $AdS_{4}$ will be incompatible with Minkowski-space supersymmetry, so that   
the uplift transforms AdS-supersymmetric mass splittings into non-supersymmetric Minkowski mass splittings.

Second, the uplift 
induces small shifts of the vevs; in particular, the shift of the (lowest component of the) K\"ahler modulus superfield results in  a nonvanishing vev for the corresponding F-term $F_{T}$.
Then, nonperturbative superpotential couplings between the visible sector and the Kähler modulus $T$ of the form (\ref{ovisis})
give rise to soft masses in the visible sector of the form
\begin{equation}
\delta {\cal{L}}_{soft} = - a\,  {\cal{O}}_{vis}\, e^{-aT} F_{T} + c.c.\,.
\end{equation}
Finally, direct communication through the bulk can induce non-holomorphic masses,
which in the four-dimensional theory arise from K\"ahler potential couplings of the form
\begin{equation} \label{d4}
\delta K = \int d^4 \theta \, Q^{\dagger}Q X^{\dagger}X \,,
\end{equation} with $X$ a spurion for hidden-sector supersymmetry breaking.
However,  as we have reviewed, \cite{KMS} demonstrated that couplings of this third type are suppressed in {\it{warped}} backgrounds, by the gravity dual of conformal sequestering.

In our explicit example, the leading contribution arose from soft masses of the first kind, i.e.\ AdS-supersymmetric mass splittings transposed to Minkowski space.  Soft masses of the second kind, i.e.\ soft masses arising from F-terms induced by uplifting, turned out to be subleading in $1/(aU), 1/B$, cf.\ the discussion surrounding (\ref{small}), but we expect that in more general models these masses will not be negligible.  We now turn to applying these considerations to a more realistic model.


\section{Soft Terms in Realistic Models}\label{sec:realistic}

The results of the preceding sections have shown that in a simple string theory toy model for the visible sector --- a single D3-brane --- soft masses of order $m_{3/2}$ are induced by superpotential cross-couplings.
This model differs from the MSSM in two very important respects. First, the `visible sector' scalar superfields, i.e.~the superfields whose lowest components are the D3-brane coordinates $z_{i}$, appeared in a rather complicated way in the superpotential (\ref{d3w}).  Of course, in the MSSM, gauge invariance and R-symmetry severely restrict the
form of the superpotential, and the only allowed terms up to cubic order in the visible fields are those in (\ref{ovisis}).
As we will see, the structure in \eqref{ovisis} provides the squarks and sleptons with some degree of protection from superpotential de-sequestering, but the Higgs sector enjoys no such protection.

Second, while some visible sector fields in the model in \S\ref{sec:toy} obtained significant vacuum expectation values and F-terms, this is not a desired feature of a realistic visible sector (before electroweak symmetry breaking). In the models that we will consider subsequently, the vanishing of the visible sector F-terms will be an assumption, while in a complete string theory construction it should be the outcome of a computation.

In this section we instead assess the effect of superpotential cross-couplings for  MSSM-like models that we assume are embedded in a string compactification through an otherwise unspecified D-brane construction. The details of the  moduli stabilization scenario have important consequences for the resulting soft terms. In \S\ref{Sec:kklt}  we study the KKLT model \cite{Kachru:2003aw}, assuming that the
supersymmetry-breaking sector is localized far down a warped throat, so that the K\"ahler potential is of the form \eqref{kseq}, while in
 \S\ref{sec:LVS} we study the Large Volume Scenario (LVS), where extended no-scale has been argued to imply a special form of the metric on the visible sector moduli space, as discussed in \S\ref{s:extns}. In both cases, soft masses can be computed within a general framework that we now outline.

Along the lines of \cite{KL}, let us consider a supergravity theory containing visible sector fields $C^{a}$ and a modulus field $T$.
(We will consider multiple moduli fields later.) In an expansion around zero vacuum expectation values
for the visible sector scalars, the superpotential and K\"ahler potential can be expanded as
\be
W & = &
\widehat{W}(\X) + \mu(\X) H_u H_d + \frac{1}{6} Y_{a
  b c}(\X) C^{a} C^{b} C^{c} + \ldots\,,\\
\label{MatterK} K & = & \widehat{K}(\X, \bar{\X}) +
\tilde{K}_{a \bar{b}} (\X, \bar{\X}) C^{a}
\bar{C}^{\bar{b}}+ \ldots\,, 
\ee
where we are assuming that no holomorphic or anti-holomorphic terms such as $Z(\X,\bar{\X}) H_{u}H_{d}+h.c.$ are present\footnote{The effects of $Z \neq 0$ are discussed e.g.~in \cite{Brignole:1997dp}. }.

As discussed in \S\ref{softsection}, a supersymmetry-breaking Minkowski vacuum is obtained by adding an additional  uplift contribution to the scalar potential. This uplift potential is in general dependent on both the visible sector fields and the K\"ahler moduli.

Then in order to compute the soft scalar masses \cite{Girardello:1981wz} (not yet canonically normalized, which we indicate with the hats) in \eqref{Lsoft}
one has to expand the total potential $V_{tot} = V_{F}+V_{up}$ (where the standard supergravity F-term potential $V_{F}$ is given in \eqref{pot}), 
\be
 B \hat{\mu}&=&\nabla_{h_u}\nabla_{h_d}(V_{F}+V_{up})\big|_{C=0}\,,\\
 \hat M^{2}_{a \bar{b}}&=&\nabla_{a}\nabla_{\bar b}(V_{F}+V_{up})\big|_{C=0}-m_{ab}K^{b\bar c}m_{\bar c \bar b}\big|_{C=0}\,,\\
 \hat A_{a b c} & = &  \nabla_{a}\nabla_{b}\nabla_{c} (V_{F} +V_{up})\big|_{C=0}\,.
\ee
Here $m_{ab}$ denotes the fermion masses, which in general are given by equation \eqref{FermGen} and are not corrected by the uplift potential. The contributions to the un-normalized soft terms from $V_F$ were obtained a long time ago
\cite{KL,Brignole:1997dp,Soni:1983rm}, 
and in this notation can be written as
\be
\nabla_{h_u}\nabla_{h_d}V_{F} & = & e^{\widehat{K}/2 \Ms} \mu \left[ F^m
\left( \partial_m \log \mu + \frac{\hat{K}_m}{\Ms}  -
\Gamma^{h_u}_{m h_u} - \Gamma^{h_d}_{m h_d}
   \right) - \frac{ \widehat{ \overline W} }{|\widehat W|}m_{3/2} \right]\,, \label{Bsoft} \\
 \nabla_{a}\nabla_{\bar b}V_{F}&-&m_{ab}K^{b\bar c}m_{\bar c \bar b}
=   \left( \frac{F^{m} K_{m \bar n} \bar{F}^{\bar n} }{\Ms} - 2 m_{3/2}^2 \right)  \tilde{K}_{a \bar{b}}\nonumber \\
&&\quad \quad - F^{\bar{m}}F^n \left(  \partial_n \bar{\partial}_{\bar m} \tilde K_{a \bar b} - \Gamma^{\bar d}_{\bar{m} \bar{a}} K_{c \bar{d}} \Gamma^{c}_{n b}
 \right) \  , \label{Msoft} \\
 \label{Aterm} \nabla_{a}\nabla_{b}\nabla_{c} V_{F} & = & e^{\widehat K/2\Ms} F^m 
  \left[ \partial_m
  Y_{a b c} +
\frac{\widehat{K}_m}{\Ms} Y_{a b c}  - \left( \Gamma^{d}_{m a}\ Y_{d b c} + (a \leftrightarrow b ) + (a \leftrightarrow c) \right) \right]\,,\,\,\,\,\, 
\ee
where every expression is to be evaluated at $C=0$ and we have specialized to Minkowski space.
The gravitino mass is given by $m_{3/2}=\frac{|W|}{\Ms} e^{K/2 \Ms}$ and we have also defined\footnote{With this definition of $F^{m}$, care is needed in lowering and raising indices, since $F_{m}\equiv D_{m}W=e^{-K/2\Ms}K_{\bar{n}m}\bar{F}^{\bar n}$.}  $F^m = e^{K/2\Ms} K^{m \bar n} {\bar D}_{\bar n} {\overline W}$ for $m, n$ taking values over all relevant moduli fields. 
Notice that corrections to the $\mu$ and $Y$ terms in the superpotential (\ref{wis}) lead to corrections in the holomorphic quadratic and cubic terms
\eqref{Bsoft} and \eqref{Aterm}, but not in \eqref{Msoft}. 

Turning our attention to the uplift potential, we will assume, as we did in \S\ref{softsection}, a term of the form
\be
V_{\rm up} =  D e^{\frac{2 K}{3M_{Pl}^{2}} } =\frac{D}{U^{2}}\,,
\ee
where $D$ is a constant, corresponding for example to supersymmetry breaking by an anti-D3-brane.
The vanishing of the visible sector vevs together with the absence of visible sector gauge-invariant linear terms in the K\"ahler potential implies that $K_{a} = 0$ (note that this is quite different from the toy model we considered 
earlier), so the only contribution to the soft terms from the uplift potential is to the soft scalar masses:
\be
\nabla_{a}\nabla_{b}\nabla_{c} V_{up} &=& \nabla_{a}\nabla_{b} V_{up}=0 \,,\label{Bup}\\
\nabla_{a}\nabla_{\bar b} V_{up} &=& \frac{2 \tilde K_{a\bar b}}{3 M_{Pl}^{2}}V_{up}\,. \label{VupM}
\ee
In order to proceed further we need to specify a scenario for K\"ahler moduli stabilization.


\subsection{KKLT stabilization \label{Sec:kklt}}

Soft terms in the KKLT scenario have been previously studied by a number of authors \cite{Choi:2005ge,Choi:2005uz,Choi:2008hn}. In this section we investigate the effects of \buzz for a MSSM-like visible sector. To be concrete,  in a theory with one volume modulus $T$ and  visible sector chiral superfields $C^{a} \equiv Q^i$, $u^i$, $d^{i}$, $L^{i}$, $e^{i}$, $H_u$, $H_d$, the separability of the $f$
function (which can be justified e.g. by warped sequestering \cite{KMS}),
implies that the lowest-dimension  terms in the Kähler potential are (again omitting purely holomorphic terms, as in (\ref{MatterK}))
\begin{align}
K &= - 3 \Ms \ln \left( T + \bar{T} - \frac{1}{3 \Ms} \sum_a C^a \bar{C}^a\right) =  - 3 \Ms \ln U. \label{RealK}
\end{align}
The Peccei-Quinn symmetry of the axion field which is the imaginary part of $T$
is only broken by nonperturbative effects, so the leading superpotential terms consistent with Standard Model gauge symmetry and with the PQ symmetry are
\be
W &=& W_0 + W^{\rm tree} + \nonumber \\
&+& {\cal{A}}_0 \Bigl(1 + \frac{\hat{\mu}}{\Ms} H_u H_d  + \frac{\lambda_{ij}^{u}}{\M^3} Q^i u^j  H_u +\frac{\lambda_{ij}^{d}}{\M^3} Q^i d^j  H_d+ \frac{\lambda_{ij}^{l}}{\M^3} L^i e^j  H_d  \Bigr) e^{-a T} \,.  \label{WSm0}
\ee
The detailed structure of the tree-level superpotential, 
\be
W^{\rm tree}(C)= \lambda^{u, \rm tree}_{i j}  Q^i \q^j  H_u + \lambda^{d, \rm tree}_{i j}  Q^i d^j  H_d + \lambda^{l, \rm tree}_{i j}  L^i e^j  H_d + W^{\rm tree}(H_u, H_d)\,,
\ee
is  dependent on the exact realization of the visible sector,
but it is generally true that $W^{\rm tree}$ does not depend on the modulus  $T$. 
We also remark that ${\cal{A}}_0$ may in general depend on additional hidden sector fields,
but for our purposes it suffices to treat ${\cal{A}}_0$ as a constant.
Let us consider in turn the various soft terms.


\subsubsection{$B\mu$ term}

The leading contribution to the $B\mu$ term comes from the F-term potential because of \eqref{Bup}. Since $K$ is of the sequestered form \eqref{RealK}, only the first and the last terms in \eqref{Bsoft} contribute, to leading order in $1/U$. The first term is of the form expected from global supersymmetry as the $T$ modulus obtains a non-vanishing F-term vev, while the last term, which is proportional to the gravitino mass, can be interpreted in KKLT as an uplifted, previously AdS-supersymmetric $B$ term.
In a perturbative expansion around the supersymmetric solution, we find
 $
F_T \approx \delta X^T \partial^2_{TT} W \axs = - \frac{3 a W}{U}\delta X^T \,,
$
where $\delta X^T$ is the shift in the real part of the volume modulus due to the uplift.
Since $K^{T \bar T} = \frac{U^2}{3 \Ms}$, we find that
\be
F^T = e^{K/2 \Ms} K^{T \bar T} \bar{F}_{\bar{T}} =  - \frac{3 a }{U} \frac{U^2}{3 \Ms}  \delta X^T e^{K/2 \Ms} \overline W = - aU \frac{\overline W}{|W|} m_{3/2}\ \delta X^T.
\ee
The leading contribution to the shift in the vacuum expectation value of $T$ is given by $\delta T = - \partial_N V_{up}  ( \partial^2_{TN} V_{tot}  )^{-1}$, which in the case of vanishing visible sector vevs
evaluates to $\delta T = \frac{2}{a^2 U}$ for the dimensionless modulus. It follows that the dimension one F-term $F^T $ 
is
\be
F^T  =   - \frac{2}{a^2U} aU \frac{\overline W}{|W|} m_{3/2} = -\frac{2}{a} \frac{\overline W}{|W|}  m_{3/2}. \label{KKLTF}
\ee
Neglecting a possible contribution to the $\mu$ term from the tree-level superpotential\footnote{A tree-level $\mu$ term of ${\cal O}(v)$ would result in corrections to this estimate of the order of $\frac{v}{U^{1/2} m_{3/2}}$. } $W^{\rm tree}$, we have
$\partial_T \log \mu(T) = -a$, and the un-normalized $B \mu$ term, $B \hat{\mu}$, becomes
\be
  B \hat{\mu} & \simeq &
e^{\hat{K}/2 \Ms} \mu \left( F^m
 \partial_m \log \mu   - m_{3/2}\frac{\overline W}{|W|} \right) = e^{\hat{K}/2 \Ms} \mu\  m_{3/2} \frac{\overline W}{|W|}\nonumber \\
 &=&- m_{3/2}^{2}\frac{3\hat \mu}{aU}. \label{KKLTBmu}
 \ee
After canonically normalizing the Higgs field kinetic terms, the physical $B\mu$ term becomes
\be
  B {\mu} & \simeq & -\frac{3 \hat{\mu}}{a} \mgrs.
\ee
We emphasize that this $\hat{\mu}$ is the coefficient in
the nonperturbative contribution \ref{WSm0}.


\subsubsection{Soft non-holomorphic masses}

The soft masses receive contributions both from the F-term potential \eqref{Msoft} and from the uplift potential \eqref{VupM}.
If the uplift had been achieved purely through the F-term vevs of the moduli fields, so that $ F^m \bar{F}^{\bar n} K_{m \bar{n}}  = 3 \mgrs \Ms$,
the first line of the right-hand side of \eqref{Msoft} could have been written simply as $+ m_{3/2}^{2}\tilde K_{a\bar b}$ in Minkowski space. In KKLT, however, the uplift is obtained by adding an explicit uplift potential, so for the F-terms in equation \eqref{KKLTF} we instead find
\be
F^{T}K_{T \bar T} \bar{F}^{\bar T} = \frac{4}{(aU)^2} \cdot 3 \mgrs \Ms.
\ee
These terms are then a priori subleading with respect to the contribution proportional to \mgrs. However, after adding the contribution from the uplift potential as in \eqref{VupM},
\be
\nabla_{a}\nabla_{\bar b}V_{up}=\frac{2 V_{up}}{3 M_{Pl}^{2}}  \tilde K_{a\bar b} = 2 \mgrs  \tilde K_{a\bar b}  \,,
\ee
the leading contribution to the sfermion and Higgs un-normalized soft masses cancels precisely. Retaining the first subleading term, one finds
\be
 \hat M^{2}_{a \bar{b}}&\simeq&\left[ 2-2 +\frac{8}{(aU)^{2}} \right] m_{3/2}^2  \tilde K_{a\bar b}   \,.
\ee
The leading-order contribution shows an exact cancellation between two terms of very different physical origin. While the negative contribution to the soft masses can be interpreted as mass splittings that were supersymmetric in AdS, the positive contribution is a direct effect of the supersymmetry-breaking uplift potential. The exact cancellation  is a consequence of the fact that the uplift potential for an anti-D3 brane is given by $\frac{D}{U^n}$ for $n = 2$.\footnote{The uplifting was described in terms of a spurion superfield in \cite{Choi:2005ge}, where the cancellation for $n=2$ was also noticed.   An off-shell formulation of this sort is not required for our purposes.}
For $n \neq 2$ --- as can be expected for uplifts corresponding to objects extended in the internal dimensions --- the soft masses would be of the order of $m_{3/2}$.  In conclusion, for the K\"ahler potential in \eqref{RealK}, the superpotential in \eqref{WSm0} and for anti-D3-brane uplifting, we find that the total soft mass for the normalized visible sector fields can be written as
\be
M^2_{a \bar b} = \frac{8}{(aU)^2} \mgrs\ \delta_{a \bar b}\,, 
\ee
which is flavor-universal and does not induce additional flavor-changing neutral currents.


\subsubsection{$A$ terms}\label{s:A}

Finally, let us consider the soft trilinear $A$  terms. We first note that $K_T = - 3 \Ms/U$,
$\Gamma^c_{T a} = - \delta^c_a \frac{1}{U}$, and  the Yukawa coupling can be written as
$Y_{u  i j} = {\cal{A}}_{0}/\M^3 e^{-aT}\lambda^u_{i j} + \lambda^{u, \rm tree}_{i j}$ for the up-type quark superfields and similarly for the down-type quarks and for the leptons. Using this in the expression
\eqref{Aterm}
for  the un-normalized $A$ terms  results in an exact cancellation of everything but the first term, as expected from the extended no-scale discussion in \S\ref{s:extns}. Thus, the un-normalized scalar trilinear $A$ term is given by
\be \label{del}
 \hat A_{uij}&=&e^{\hat K/2\Ms} F^m   \partial_m  Y_{uij} \nonumber \\
&=& -\frac6{aU}\ \frac{m_{3/2}}{ M_{Pl}} m_{3/2}\    \lambda^u_{i j}\,.
\ee
After electroweak symmetry breaking $ \langle H_{u,d}^{0} \rangle\equiv v_{u,d}$, the $A$ terms will contribute to the sfermion masses. Unless $A$ is  real and
has very special flavor structure,
this will induce FCNC and CP violation. A useful quantity for which model-independent experimental constraints are available \cite{Gabbiani:1996hi} is $\delta\equiv \Delta_{a b}/M_{\rm soft}^{2}$, where $\Delta_{a b}$ is the flavor-off-diagonal contribution to the propagator of the sfermions in the basis in which the couplings to neutral gauginos are flavor diagonal, and $M_{\rm soft}$ is the average sfermion mass. In principle the soft terms we compute are valid at some high supersymmetry breaking scale and should be run down to the scales where experiments are performed using RG equations. The largest running appears in the third-generation quarks because of the large top Yukawa coupling. Since the experimental data
we focus on does not involve the third generation, we ignore the effects of RG running.

E.g. for canonically normalized up-type squarks  
 we find
\be \label{delt}
\delta\equiv \frac{\Delta_{a b}}{M_{\rm soft}^{2}}&=& -\frac{6 U^{1/2} }{a}\ \left(\frac{m_{3/2} }{\M}\right) \left(\frac{m_{3/2} v_u }{  M_{\rm soft}^{2}   }\right) \    \lambda^u_{i j} \,. 
\ee
For a numerical estimate, as in \S\ref{BoundBSect}, we take $m_{3/2}\sim aU M_{\rm soft}$ \cite{Choi:2005ge,Choi:2005uz}, $aU\sim 66$,  $a = \frac{2 \pi}{32}$ and $|\lambda_{a b}|\sim1$. We find
\be
\delta^{u,d} \sim 10^{-11} \frac{v_{u,d}}{100 \rm GeV}\,,
\ee
which is at least four orders of magnitude away from any experimental constraint from FCNC and CP violation \cite{Gabbiani:1996hi}.


\subsubsection{Summary of KKLT phenomenology}

In sum, in the KKLT scenario, flavor diagonal  squark masses of the order of $\frac{1}{(aU)} m_{3/2}$ are induced, resulting in negligible flavor violation \cite{Choi:2005ge,Choi:2005uz}. Superpotential de-sequestering in the Higgs sector can lead to a significant $B\mu$ term of order $m_{3/2}^2$, and to supersymmetric masses of order $m_{3/2}$,
which can create serious problems for electroweak symmetry breaking.

We note that as the vacuum expectation value of 
$F_T$ becomes important for the soft terms, sequestering breaks down severely, in that modifications to the compactification located far from the visible sector in the internal dimensions
can have a significant impact on $F_T$, which in turn affects the physics of the visible sector. For instance, the value of $F_T$ in \S\ref{BoundBSect} where the effects of a nearby D3-brane  
are taken into account is exactly half of that 
for the KKLT scenario in the absence of the D3-brane, cf.\ the discussion around equation \eqref{FT}. Regarding the D3-brane as a part of the hidden sector, we note that this reduction in 
$F_T$ leads to a cancellation of the leading-order $B\mu$ term \eqref{KKLTBmu}. Instead of being
of order $\mgrs$, the $B\mu$ term will in this case enter at ${\cal O}(\mgrs/U) $. A similar sensitivity to the global compactification --- though perhaps not as striking ---  is a general feature of the contributions proportional to
$F_T$ that are induced by superpotential de-sequestering.

%
%
%


\subsection{The Large Volume Scenario  \label{sec:LVS}}
So far we have only discussed the effects of \buzz in KKLT compactifications, but we expect significant effects whenever a nonperturbative superpotential plays an important role in the stabilization of the moduli. We illustrate this point by turning to the Large Volume Scenario (LVS), one of the most promising areas for successful phenomenology  from string compactifications \cite{BBCQ}. In LVS, $\alpha'$ corrections to the Kähler potential are included in addition to the the same nonperturbative superpotential as in KKLT \eqref{WSm0}, both effects breaking no-scale. For Calabi-Yau manifolds of the Swiss-cheese type,\footnote{To avoid cluttering our notation with another index we write formulae for a single $T_{s}$. The generalization to many $T_{s}$'s is straightforward.  Moreover, adding a vanishing four-cycle $T_{a}$ as in \cite{Blumenhagen:2009gk,Blumenhagen:2007sm} would not change our discussion.}
\be
{\cal V} \simeq \tau_{b}^{3/2} - \tau_{s}^{3/2} \,,
\ee
a non-supersymmetric AdS vacuum exists where the overall volume ${\cal{V}}$ is exponentially large, $\V\simeq \tau_{b}^{3/2}\simeq e^{a\tau_{s}}$, with $\tau\equiv (T+\bar{T})/2$. Hence in $W$ only the nonperturbative terms for the small volumes $\tau_{s}$ are relevant. So we have
\be
W &=& W_0 + W^{\rm tree} + \nonumber \\
&+& {\cal{A}}_0 \Bigl(1 + \frac{\hat{\mu}}{\Ms} H_u H_d  + \frac{\lambda_{ij}^{u}}{\M^3} Q^i u^j  H_u +\frac{\lambda_{ij}^{d}}{\M^3} Q^i d^j  H_d+ \frac{\lambda_{ij}^{l}}{\M^3} L^i e^j  H_d  \Bigr) e^{-a T_{s}} \,.  \label{WSm}
\ee
It is important to notice that LVS crucially requires at least two K\"ahler moduli, in which case, as opposed to the case of a single overall volume, the separability (\ref{fseq}) of $f$ into hidden and visible sector contributions is far from established. Despite this fact, the milder condition of extended no-scale discussed in \S\ref{s:extns} has been argued to hold at least to leading order in four-cycle volumes. Let us review the argument given in \cite{Conlon:2006tj}.
Recall that for a diagonal visible sector matter metric, 
$\tilde{K}_{a \bar{b}} = \tilde{K}_a \delta_{a \bar{b}} \  {\rm (no\ sum)}$, the physical Yukawa couplings in supergravity are given by
 \be
 Y^{\rm phys}_{abc} = e^{K/2\Ms} \frac{Y^{\rm hol.}_{abc}}{\sqrt{K_a K_b K_c}} \  ,
 \ee
and can be computed from first principles in a given localized model by computing the overlap of the corresponding normalized wavefunctions.
In \cite{Conlon:2006tj}, it was argued that locality implies that  the physical Yukawa couplings  should be independent of the volume modulus to all orders in perturbation theory.
Since the \emph{holomorphic} Yukawa couplings are independent of the volume modulus to all orders in perturbation theory by holomorphy combined with the Peccei-Quinn symmetry,
this requirement enforces a special form of the metric on the visible sector moduli space: 
$\tilde{K}_a \sim e^{\widehat{K}/3\Ms}$. This leads to the structure studied in \S\ref{s:extns} \cite{Blumenhagen:2009gk},
\be
\tilde{K}_{a \bar{b}}=e^{\widehat{K}/3 \Ms} \kappa_{a\bar b}\,,
\ee
with $F^{m}\partial_{m} \kappa_{a\bar b}=0$, i.e.~$\kappa_{a\bar b}$ does not depend on the moduli that get non-vanishing F-term vevs. 

Superpotential cross-couplings induce corrections to the $B\mu$ terms and $A$ terms but not to the nonholomorphic sfermion masses $M^{2}$. Given \eqref{Bup}, the corrections $\delta B\mu$ and $\delta A$ can be computed from the first term on the right hand side of \eqref{Bsoft} and \eqref{Aterm}, respectively. We now consider  these contributions in turn.


\subsubsection{$B\mu$ term}

We want to compute the contribution to the $B\mu$ term from superpotential cross-couplings, which is given by the first term in \eqref{Bsoft}. To canonically normalize the visible sector fields we need $\tilde K_{a\bar b}$. For our purposes, it is sufficient to know it at leading order in the large volume expansion, since there are no precise cancellations that make the subleading orders important. We can hence use the extended no-scale relation \cite{Conlon:2006tj} and rotate and rescale the fields to have
$\tilde K_{a\bar b}=e^{\widehat{K}/3M_{Pl}^2}\delta_{a\bar b}$.
 Then we find the correction from superpotential cross-couplings to the canonically normalized $B\mu$ term to be
\be
\delta B\mu&=& e^{\widehat K/6\Ms} F^m   \partial_m \mu  \\
&=&-\frac{a}{{\cal V}^{1/3}}F^{T_{s}}e^{-a\tau_{s}}\frac{{\cal{A}}_0 \hat{\mu} }{M_{Pl}^{2}}\,.
\ee
For an estimate, we assume $|{\cal{A}}_0|\sim M_{Pl}^{3}$, $|\hat \mu|\sim 1$, 
and use the
facts that in LVS,
the
volume at the minimum of the potential
is ${\cal V}\simeq e^{a \tau_{s}}$ and $|F^{T_{s}}|\simeq m_{3/2}$. Keeping track only of volume factors we find
\be \label{BLVS}
\delta B\mu\sim \frac{M_{Pl}^{2}}{{\cal V}^{7/3}}\,.
\ee
For
successful electroweak symmetry breaking, $B\mu$ should not be far from the weak scale. Hence, unless this contribution is absent, \eqref{BLVS} puts a strong lower bound on the size of the overall volume: ${\cal V}\gtrsim 10^{14}$ in string units.


\subsubsection{$A$ terms}

Now let us turn to the contribution to
the soft $A$ terms from superpotential cross-couplings. This is given by the first term in \eqref{Aterm}. After canonically normalizing the visible sector fields using 
$\tilde K_{a\bar b}=e^{\widehat{K}/3M_{Pl}^2}\delta_{a\bar b}$ \cite{Conlon:2006tj} at leading order (since again there is no relevant cancellation for the terms we are computing), we have
\be
\delta A_{uij}&=&e^{-\widehat K/2\Ms}e^{\widehat K/2\Ms} F^m   \partial_m  Y_{uij} \nonumber \\
&=&-\frac{{\cal{A}}_0 \lambda_{ij}^{u} }{M_{Pl}^{3}} a F^{T_{s}} e^{-a\tau_{s}}\,,
\ee
and similarly for $\delta A_{dij}$ and $\delta A_{lij}$. As we saw in \S\ref{s:A}, an efficient way to estimate the phenomenological effect of this correction,\footnote{Again, for the quantities of interest 
here
we expect
the RG evolution to give only negligible corrections.} e.g.\
in terms of FCNC and CP violation, is to compute the parameter $\delta$ defined in \eqref{delt}. Using
$e^{a\tau_{s}}\simeq {\cal V}$ as above, we write
\be
\delta=\frac{{\cal{A}}_0 \lambda_{ij} }{M_{Pl}^{3}}\frac{a}{{\cal{V}}}\frac{v F^{T_{s}}}{M^{2}_{\rm soft}}\,.
\ee
Let us focus just on the scaling with the overall volume, neglecting factors of $a$
and
$T_{s}$. Again we take
$|{\cal{A}}_{0}|\sim M_{Pl}^{3}$, $|\lambda_{ij}|\sim 1$ and $|F^{T_{s}}|\simeq m_{3/2}$. The size of the last ingredient, $M_{\rm soft}$, being the average value of the relevant soft terms, is still a matter of debate \cite{Choi:2010gm,Conlon:2005ki,Blumenhagen:2009gk}, so  we parameterize it as
\be
M_{\rm soft}^{2}\sim \frac{ m_{3/2}^{2}}{{\cal{V}}^{n}}\,,
\ee
where $n$ has been claimed to be $\{0,1/3,1\, \mathrm{or}\, 2\}$ in \cite{Choi:2010gm,Conlon:2005ki,Blumenhagen:2009gk} respectively. Putting things together we find
\be
\delta\sim {\cal V}^{n} 10^{-16} \left(\frac{v}{100\,\rm GeV}\right)\,.
\ee
One of the strongest experimental constraints on $\delta$ comes from bounds on $\mu\rightarrow e\gamma$ and gives $|(\delta^{l}_{12})_{LR}|<2\cdot10^{-6}$ \cite{Gabbiani:1996hi}. Other strong constraints arise due to the CP violation induced by ${\rm Im}\,\delta$, e.g.~from electric dipole moments. Figures of merit are $|{\rm Im}(\delta^{d}_{11})_{LR}|<3 \cdot 10^{-6}$, $|{\rm Im}(\delta^{u}_{11})_{LR}|<6 \cdot 10^{-6}$ and $|{\rm Im}(\delta^{l}_{11})_{LR}|<4 \cdot 10^{-7}$.

Unless the nonperturbative correction is real and respects flavor (more on this point in \S\ref{Sec:Corrs}), we find the following upper bound on the size of the overall volume in LVS:
\be \label{volumebound}
{\cal V}<10^{10/n}\left(\frac{v}{100\,\rm GeV}\right)^{1/n}\,.
\ee 
For $n=0,1/3$, as claimed in \cite{Choi:2010gm,Conlon:2005ki}, respectively, this bound is irrelevant. On the other hand, in the case considered in \cite{Blumenhagen:2009gk} one finds the upper bounds $\V<10^{10}$ and $\V<10^{5}$ for $n=1$ and $n=2$, respectively. Since in \cite{Blumenhagen:2009gk} it was argued that the smallest possible volume is $\V\sim 10^{6-7}$ in string units, for $n=2$ there might already be some tension with experiments due to
superpotential de-sequestering. A careful analysis keeping track of all $\mathcal{O}(10)$ factors neglected in the above estimate would be desirable. Finally, notice that this bound is in contradiction with the one obtained from the size of the $B\mu$ term \eqref{BLVS}, leading to an inconsistency unless one of the two contributions is forbidden by some additional mechanism.
Summarizing, the bound \eqref{volumebound} shows that data on CP violation and FCNC could be used to rule out an interesting region of parameter space.


\subsection{On nonperturbative corrections to visible sector superpotentials.} \label{Sec:Corrs}

Our discussion so far has assumed that because nonperturbative superpotential couplings between the visible sector and the K\"ahler moduli are not forbidden by known symmetries, these couplings are in fact present, and are not simply proportional to the tree-level Yukawa couplings. That is, if the visible-sector superpotential includes terms of the form\footnote{To simplify our expressions, in this section we will focus on u-type quarks; the extension to the remaining fermions is trivial.}
\begin{equation}
W_{vis} 
= \lambda_{ij}^{u, {\rm tree}} Q^i u^j H_{u} + \lambda_{ij}^u Q^i u^j H_{u} \,e^{-a T} \,,
\end{equation}
then our working assumption has been that $\lambda_{ij}^{u, {\rm tree}}$ and $\lambda_{ij}^{u}$ are not proportional.

This assumption is strongly supported by the abundance of examples in the literature in which string instantons or D-brane instantons give rise to Yukawa couplings that are forbidden in perturbation theory (see \cite{Mirjam} for a comprehensive review in the context of intersecting D6-branes in type IIA orientifolds, and \cite{Buican:2008qe} for a discussion of D-brane instantons stretched between the visible and hidden sectors.)  In F-theory GUT models, contributions from nonperturbative effects on D7-branes have been argued to solve the `rank problem' of the tree-level Yukawa couplings \cite{Marchesano:2009rz}, which obviously requires an adjustment of the flavor structure.
Thus, flavor violation by nonperturbative effects is well-attested in string-theoretic realizations of the MSSM.

It would be interesting to obtain further details on the form of nonperturbative superpotential couplings by direct computation in string models.  A full treatment of this point is beyond the scope of the present work,
but we now outline what needs to be done to evaluate this.

%

\subsubsection{$A$ terms}


To acquire a more detailed picture of nonperturbative contributions to $A$ terms in type IIB compactifications, we now examine
an analogous computation for adjoint open string fields \cite{Berg:2004ek}.
There,
the moduli-dependent
string S-matrix of  D7-brane gauge field vertex operators
$\V_{A_{\mu}}$
was considered:\footnote{To be precise, the string S-matrix is only obtained as an explicit function
when the worldsheet moduli are integrated over, as in \cite{Berg:2004ek}.}
\be
\langle \V_{A_{\mu}}\V_{A_{\nu}}\rangle_{\mbox{\small background $\phi$ }}    = \mbox{  function of moduli } S,T,U,\phi \; .
\ee
From this string correction to the physical gauge couplings one extracts
a correction to the holomorphic gauge kinetic function
of gauge fields on D7-branes:
\be
 f_{\rm 1-loop} = -2\ln \tht_1(\phi,U) + \ldots
\ee
where the omitted terms are independent of $\phi$ and $\tht_1(\phi,U)$ is a Jacobi theta
function.
Now, substituting $ f_{\rm 1-loop} $ in the nonperturbative superpotential
on D7-branes, 
\be
W = {\cal{A}} e^{-a(f_{\rm tree} + f_{\rm 1-loop})}\,, 
\ee
we obtain
\be
W = {\cal{A}}(\phi) e^{-aT}\,, 
\ee
where
\be
{\cal{A}}(\phi) = (\tht_1(\phi,U))^{2a} \; .
\ee
This is a toy model, but it was shown in
\cite{BDKMMM} that
analogous expressions appear in the backgrounds of interest. The key point
here is simply that ${\cal A}$ inevitably depends on $\phi$,
and the dependence is not in any way negligible.

\begin{figure}
\begin{center}
\includegraphics[width=11cm]{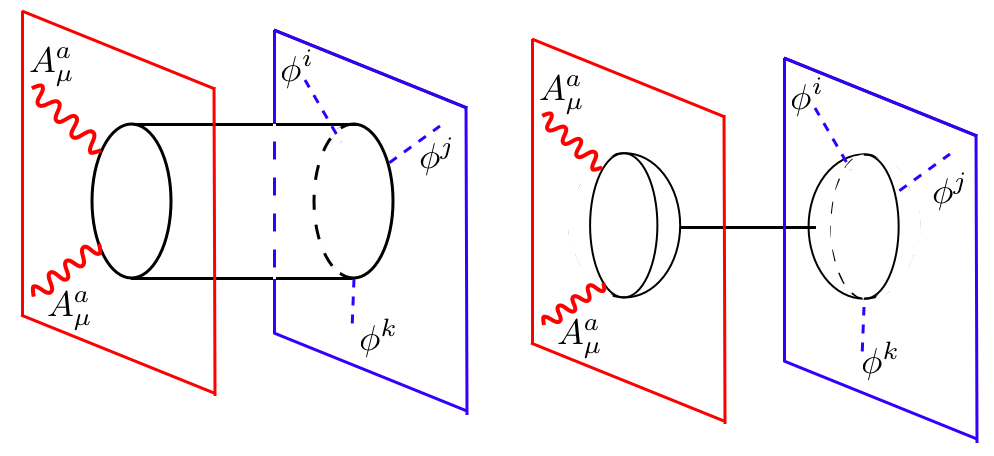}
\caption{D7-branes on the left of cylinder (in red), D3-branes
on the right of cylinder (in blue).
Left panel:
Five-point function
 analogous
to the two-point function considered in \cite{Berg:2004ek}.
Right panel:
the factorized limit. The flavor structure of $\lambda_{ij}^{u}$ may
in principle be
determined by that of the tree-level (disk) three-point function. }
\label{diagfig}
\end{center}
\end{figure}

We expect a similar calculation to  be feasible also for chiral matter fields, though more challenging.
In that case, one would expand the function ${\cal{A}}$ in gauge-invariant operators,\footnote{A
variation on this is quite common in D-brane models:
$T$ may be charged under an anomalous $U(1)$,
so that $A(\phi)$  must also transform
under the $U(1)$.  To exclude this possibility, we will assume any anomalous $U(1)$'s are broken near the string scale,
as is often the case. }
\be
{\cal{A}}(\phi^i) = {\cal{A}}_0 + {\cal{A}}_{ij}^{u} Q^i u^j H_{u} + \ldots \,, \label{Aexp}
\ee
where it is understood that these are small fluctuations
around the final, nonsupersymmetric, minimum.\footnote{Another example of charged fields analogous to \eqref{Aexp} is \cite{Haack:2006cy}.}
We cannot turn on a background for chiral fields,
so this coupling needs to be probed by the  S-matrix
due to the following
five-point function:
\be
\langle {\rm tr} (\V_{A_{\mu}} \V_{A_{\nu}}) {\rm tr} (\V_{\phi_i} \V_{\phi_j} \V_{\phi_k}) \rangle\,, 
\ee
where the traces are over the respective gauge groups. This is a double trace operator and so will appear at loop level.
The vertex operators $\V_{\phi_i}$ for chiral fields
are given by the usual vertex operators for open string scalars $\phi$
but with boundary changing operators $\sigma$ that change boundary condition
from one brane stack $i$ to the next stack $j$.
The cylinder diagram one needs to compute is
shown in the left panel of figure \ref{diagfig}.
As shown in the right panel, this can factorize
onto some closed string field, call it  $X$.\footnote{Because of the PQ symmetry, $X$ cannot be $T$.} If the field $X$
appears in the moduli-dependent superpotential as ${\cal{A}}_{ij}^{u}(X)Q^i u^j H_u$
(this is sometimes described as  $X$ ``carrying flavor''),
the resulting coupling will be problematic in general, as we do not expect $\lambda_{ij}^{u, {\rm tree}} \propto \lambda_{ij}^{u}$.


However, let us mention a possible mechanism by which the nonperturbative contributions might respect the flavor symmetry preserved by the tree-level couplings.
In intersecting brane models, and for some  models with
branes at singularities (e.g. ${\cal{O}}(-3)_{\mathbb{P}^2} \sim {\mathbb{C}}^3/{\mathbb{Z}}^3$)
the Yukawa couplings arise as triple intersections
between three brane stacks. Consider the D3-brane part of the  right panel
of
figure \ref{diagfig}; without the closed string insertion $X$,
the tree-level three-point diagram
is what generates Yukawa couplings $y_{ij}$
in the first place. It is
possible that in the low-energy limit,
the coupling $\lambda_{ij}^{u}$ that is generated
could satisfy
\begin{equation} \label{MFV}
\lambda_{ij}^{u} = c\,\lambda_{ij}^{u, {\rm tree}}
\end{equation} for some constant $c$,
so that no new flavor-changing effects are induced.  However, a much more detailed investigation would be required to establish a mechanism along these lines.

We conclude that there is no evidence that the couplings in \eqref{Aexp} should vanish, nor is there
currently a compelling argument that these couplings should generically preserve the tree-level flavor structure.

\subsubsection{$\mu$ terms}

The $\mu$ term is often prohibited at a high scale by a continuous PQ symmetry that leaves a remnant discrete symmetry.  In such a case, one should ask whether $T$ is charged under the PQ symmetry: if not, then the PQ symmetry forbids nonperturbative corrections to the (prohibited) tree-level $\mu$ term.  If instead $T$ carries a PQ charge, one would have to explain why the moduli-stabilizing superpotential itself is not forbidden.  This question depends in some detail on the particular realization of the MSSM, and as such is beyond the scope of this work.

\section{Conclusions}\label{sec:conclusions}

Higher-dimensional locality provides a promising organizing principle for the suppression of the phenomenologically dangerous flavor-violating soft terms of gravity-mediated supersymmetry breaking. Although locality in barren extra dimensions leads to sequestering, moduli-mediated couplings in the K\"ahler potential prevent sequestering in generic unwarped compactifications of string theory \cite{Anisimov:2002az}, even after moduli stabilization \cite{Kachru:2006em}. Strong warping suppresses K\"ahler potential couplings \cite{KMS} via the gravity dual of conformal sequestering, but we have shown that
nonperturbative stabilization introduces new superpotential couplings between the K\"ahler moduli and the visible sector.
As analogous couplings are known to violate tree-level flavor symmetries in many examples, it is reasonable to expect flavor violation in this case as well.  

In a very simple and explicit toy model involving a D3-brane in a conifold region of a KKLT compactification, we showed that some of the soft scalar masses induced by the superpotential cross-couplings are of order the gravitino mass. Therefore, supersymmetry breaking in this model does not sequester, in contrast to the positive result obtained by \cite{KMS} for the corresponding configuration before K\"ahler moduli stabilization.

Our considerations also apply to more realistic visible sectors.
In a KKLT compactification with an MSSM-like sector assumed to be supported on a collection of D-branes separated from the supersymmetry-breaking sector by warping, there are flavor-universal contributions to the sfermion masses of the order of $m_{3/2}/(aU)$.
In the Higgs sector, nonperturbative superpotential cross-couplings induce $\mu$ and $B\mu$ terms of order $m_{3/2}$ and \mgrs, respectively, with a detailed sensitivity to the global compactification.  Thus, in KKLT compactifications, the sequestering described in \cite{KMS} does not survive moduli stabilization. However, the corrections due to \buzz are rather mild: 
depending on the full mediation scenario, the corrections to the squark and slepton masses due to \buzz can be made sub-dominant, 
while for the Higgs sector the nonperturbative superpotential contributions to
the
$\mu$ and $B\mu$ terms
 are not necessarily fatal but must be properly incorporated, as in \cite{Choi:2005uz}.

In the case of the Large Volume Scenario --- in which the supersymmetry breaking sector is in no way geometrically separated from the visible sector --- ten-dimensional locality can still result in a significant suppression of the soft masses with respect to the gravitino mass \cite{Blumenhagen:2009gk}. However, the nonperturbative superpotential is an essential ingredient in the moduli stabilization, and
we found that
in certain parameter regimes, soft trilinear $A$ terms induced by nonperturbative superpotential couplings can be dangerously large, so that precision results from flavor physics constrain the model parameters.  Intuitively, 
scenarios with large hierarchies between $m_{3/2}$ and
$M_{soft}$
are the most vulnerable to small corrections from nonperturbative superpotential cross-couplings, and indeed we found a flavor and CP problem  only in the scenario of \cite{Blumenhagen:2009gk}, where such a hierarchy is present, while non-negligible corrections to the $B\mu$ term are to be expected in essentially all scenarios.

Our work sharpens the criteria for sequestered supersymmetry breaking in a string compactification: examination of the K\"ahler potential alone is insufficient when nonperturbative superpotential terms control the stabilization of moduli.  Two additional tasks are required: one should provide a mechanism that controls nonperturbative contributions to $B\mu$, and one must ensure that
any flavor-violating nonperturbative contributions to the $A$ terms are consistent with experiment.

A very interesting task for the future is the construction of an explicit, realistic visible sector in string theory for which the soft masses are sequestered.

\section*{Acknowledgments}
We are grateful to Monika Blanke, Joseph Conlon, Nathaniel Craig, Csaba Cs\'aki, Daniel Green, Yuval Grossman, Shamit Kachru, Sven Krippendorf, Anshuman Maharana, Aaron Pierce, Kuver Sinha, and Jesse Thaler for helpful discussions.
We thank Michael Haack, Shamit Kachru, Sven Krippendorf, and Raman Sundrum for insightful comments on the manuscript.
The research of L.M. was supported by the Alfred P. Sloan Foundation and by the NSF under grant PHY-0757868.  We gratefully acknowledge support for this work by the Swedish Foundation for International Cooperation in Research and Higher Education.

\appendix

\addtocontents{toc}{\protect\setcounter{tocdepth}{1}}

\section{Smeared Sources and Warped Sequestering} \label{smearing}
In this Appendix we show that warped sequestering in the no-scale setup of \cite{KMS} survives the relaxation of an assumption made for technical simplicity in \cite{KMS}:
smearing of the supersymmetry-breaking anti-D3-brane around the $S^{3}$ tip of the warped deformed conifold is not required for the basic conclusion to hold.


DeWolfe, Kachru and Mulligan \cite{DKM} considered a D3-$\overline{\rm D3}$ pair smeared around the $S^{3}$ tip of a Klebanov-Strassler throat, and obtained a supergravity solution at large radius in which all fields were invariant under the $SU(2)_L \times SU(2)_R$ isometry.\footnote{See \cite{MSS,BGH,BGGH} for further work on supergravity solutions for antibranes in the background of \cite{Klebanov:2000hb}.}
However, it is clear that a brane placed at a particular position on the $S^{3}$ will break some of the symmetries of the system, and can be expected to source modes that are not global symmetry singlets.  One can ask whether such modes will, at the {\it{nonlinear}} level,
induce $\Phi_-$ perturbations, and hence D3-brane soft terms, that compete with those mediated by ${\cal{O}}_{8}$.  That is, one should ask whether soft masses computed in the smeared solution of \cite{DKM} are in fact the leading soft masses in a full, unsmeared solution.  We will now argue that a solution describing a single anti-D3-brane placed at the tip of the deformed conifold enjoys an $SU(2)$\
symmetry in \slsr and that this residual symmetry forbids the $\Delta=5/2$ operator \ab, which might otherwise be expected to induce problematic soft masses.\footnote{See
\cite{fluxpaper} for a setting in which nonlinear effects of this operator indeed give the dominant contribution to the D3-brane potential.}

The deformed conifold can be defined as the a subset of $\C^4$ satisfying
\be
\det W = - \frac{\epsilon^2}{2} \,, \label{W}
\ee
with
\be
W  =  \left( \begin{array}{cc} -w_3 & w_2 \\
-w_1 & w_4 \ea
 = \frac{-1}{\sqrt{2}} \ba z_3 + i z_4 & z_1 - i z_2 \\
 z_1 + i z_2 & -z_3 + i z_4 \ea . \label{Wcoords}
\ee
The radius $r$  is given by
\be
\tr W^{\dagger} W = \sum_i |z_i|^2 = r^3. \label{r}
\ee
In the matrix representation of the coordinates of a point $p$, it is easy to convince oneself  that \slsr acts transitively on the $T^{1,1}$ base of the cone as
\be
\sigma( W(p), g) \rightarrow L(g) W(r) R(g)^{\dagger}\,,  \\
g \in \slsr .
\ee
This means that we can choose an origin, $W_0$, and specify all other points by \slsr transformations away from this point. It is standard to choose
\be
W_0 = \ba \frac{\epsilon}{\sqrt{2}} & \sqrt{r^3 - \epsilon^2 } \\
0 & -\frac{\epsilon}{\sqrt{2}}\,, \ea \label{W0}     \, ,
\ee
and then any point on $T^{1,1}$ at this radius can be obtained through
\be
W = L W_0 R^{\dagger}. \label{Wrotation}
\ee
The background geometry and the smeared solution are symmetric under all these rotations. The subset of these rotations that are still symmetries once an anti-D3-brane is placed at a specific point $p$ is by definition the stabilizer $H(p)$. Since $T^{1,1}$ is a coset space of the above \slsr action, the stabilizer is (cf.\ e.g.\ \cite{Candelas}),
\be
H(p) =  \left\{
 \begin{array}{l c} U(1) & r^3 > \epsilon^2 \\
SU(2) &  r^3 = \epsilon^2\,.
  \end{array} \right.
\ee
We denote the stabilizer of our chosen origin $W_0$\  by  $SU(2)_S$, and we must impose the $SU(2)_S$\ symmetry on any solution corresponding to perturbing the supergravity solution by placing an anti-D3-brane at $p$. Clearly, for this specific $W_0$, the stabilizer is the subgroup of $SU(2)_L \times SU(2)_R$ that leaves $\sigma_3$ invariant.

Now, because $\det_{i, j} \ab = 0$, \ab  cannot be a singlet under $SU(2)_S$. Geometrically, this is the natural outcome of the fact that  $A^i B^j$\ can be
thought of as coordinates on the \emph{singular} conifold.
Thus, the operator \ab has an interpretation as a point on the conifold (far) away from the tip. Such a point has a $U(1)$ stabilizer, and thus  cannot be invariant under a  full $SU(2)$ in \slsr. We conclude that the corresponding supergravity mode will not be turned on even in an unsmeared solution.


\section{A D3-brane on the Conifold}
\label{conifold}
In this Appendix we give a detailed treatment of the toy model of \S\ref{sec:toy}, after collecting the relevant supergravity formulas in \S\ref{AppGen}.

\subsection{General strategy \label{AppGen}}
We 
first assemble some well-known expressions pertinent for the evaluation of mass matrices for general chiral superfields. As in the bulk of the paper,
we will work with the Kähler potential $K$  and the superpotential $W$  for some chiral superfields in fixed Kähler gauge.
The F-term scalar potential is
\be
V_F = e^{K/\Ms} \left( \KAB D_{A}W D_{\bar B} \overline{W} - 3{|W|^2 \over \M^2} \right)\,,
\label{pot}
\ee
where the K\"ahler covariant derivative is $D_A W = (\partial_A + \frac{K_A}{\M^2})W$.
We are interested in expressions for the masses Taylor expanded around a supersymmetric point denoted $Z_0$, for which $F_A \equiv D_A W \axs = 0$, for all values of the index $A$. The scalar mass matrices at this point are given by
\begin{align}
\nabla_a \nabla_{\bb} V_F |_{Z_0}  =&
\,  \partial_a \partial_{\bb} V_F  |_{Z_0}  = e^{K/ \Ms} \left(  \KAB \partial_a (D_A W) \partial_{\bb} (\bar{D}_{\bar{B}} \overline{W})   - 2 \frac{|W|^2}{M^4_{Pl}} K_{a \bb} \right)\,, \label{Fmassabb} \\
\nabla_a \nabla_{b} V_F  |_{Z_0} =&
\,  \partial_a \partial_{b} V_F  |_{Z_0} = - \frac{e^{K/\Ms} \overline{W}}{ \Ms} \partial_{ a } D_{ b }W, \label{Fmassab}
\end{align}
where we have used that $\partial V_F \axs = 0$ to replace the covariant derivatives with partial derivatives. The first-order corrections from the F-term potential, $\delta Z^M \nabla_M \partial^2_{ a \bb} V_F$ and $\delta Z^M \nabla_M \partial^2_{ a b} V_F$,
are obtained by taking three derivatives on the F-term potential:\footnote{Since the expressions turn out to be slightly lengthy, only the terms with two holomorphic
indices
and one anti-holomorphic index are worked out here. To obtain the complete set of corrections it is necessary to also work out the case when all indices are holomorphic.}
\begin{align}
\partial_{a b} \partial_{\bar{c}} V_F |_{Z_0}
&= \partial_{\{ a} ( e^{K/\Ms} \Kcd )\partial_{b \}} (F_c) \partial_{\bar{c}} (\bar{F}_{\bar{d}}) +  \partial_{\{ a} ( e^{K/\Ms} \Kcd )\partial_{\bar{c}} (F_c) \partial_{b \}} \bar{F}_{\bar{d}} + \nonumber \\
&+ \partial_{\{ a} (F_c) \partial_{b \}} (\bar{F}_{\bar{d}}) \partial_{\bar{c}}( e^{K/\Ms} \Kcd) + \nonumber \\
&+  e^{K/\Ms} \Kcd \left[ \partial^2_{ab} (F_c) \partial_{\bar{c}} \bar{F}_{\bar{d}} + \partial^2_{\bar{c} \{a} (F_c) \partial_{b \}} (\bar{F}_{\bar{d}}) + \right. \nonumber \\   & \left.
+ \partial_{\{ a} (F_c) \partial^2_{b \} \bar{c} } ( \bar{F}_{\bar{d}})  + \partial_{\bar{c}}(F_c) \partial^2_{ab} (\bar{F}_{\bar{d}}) \right] - 3\left[ e^{K/\Ms} \frac{W}{\Ms} \partial^2_{ab} \bar{F}_{\bar{c}}  \right]. \label{Vabc}
\end{align}

The fermion masses are
\be
m_{ab} = e^{K/2\Ms} \left[ \partial_a D_b W + \frac{K_a}{\Ms} D_b W - \Gamma^d_{ab} D_d W  \right], \label{FermGen}
\ee
where the Christoffel symbol is constructed out of the Kähler metric, $\Gamma^d_{ab} = K^{d \bar{c}} K_{a \bar{c} b}$. In an expansion around $Z_0$, we will be interested in the first-order corrections to the supersymmetric masses, which are given by
\be
m_{ab} \axst  = m_{ab} \axs + \delta Z^M (\nabla_M m_{ab}) \axs \, ,\label{FermExp}
\ee
where $M$ runs over both holomorphic and anti-holomorphic indices. The lowest-order contribution is then
\begin{align}
m_{ab} |_{Z_0} =  e^{\frac{K}{2 \Ms} }   \partial_a D_b W . \label{FermAtX0}
\end{align}
The vanishing of the F-terms implies that $\partial_a D_b W \axs = \partial_b D_a W \axs $.

The gravitino mass is given by $\mgrs = e^{K/\Ms} \left|\frac{W}{\Ms}\right|^2$.
At the supersymmetric minimum, $\mgrs \axs = \frac{|V_F|}{3 \Ms}$, and to linear order in the expansion around this minimum,
\be
\mgrs = \mgrs \Bigr|_{Z_{0}} + \delta Z^M \partial_M \left( e^{K/\Ms} \left|\frac{W}{\Ms}\right|^2 \right)\Bigr|_{Z_{0}}  = \mgrs \Bigr|_{Z_{0}}\,,
\ee
where the last step follows from  F-flatness at $Z_0$. The small mixing terms between the gravitino and the chiral fermions are proportional to $e^{K/2\Ms} \frac{F_a}{\Ms}$, and will henceforth be neglected.

Specializing to the KKLT scenario with Kähler potential $K = - 3  \Ms \ln U$ and an uplift potential of the form $V_{\rm up} = \frac{ D}{U^2} = 3 \mgrs \Ms$, 
the masses from the uplift potential can be written as
\begin{align}
\partial^2_{M N} V_{\rm up} = \frac{2 V_{\rm up}}{3 \Ms}\left( K_{MN} + \frac{2}{3 \Ms}K_M K_N \right).
\end{align}
To obtain the 
first-order correction to the mass matrix,
following the logic for corrections from $V_F$ above,
we take  three derivatives of the uplift potential,
\begin{align}
\partial^3_{MNP} V_{\rm up} = \frac{2 |V_F|}{3 \Ms} \left[ K_{MNP} + \frac{2}{3 \Ms}(K_{M} K_{NP} + {\rm cycl. perm.}) - \frac{16}{(3 \Ms)^2} K_M K_N K_P \right].  \label{UpMassCorr}
\end{align}
We will discuss the role of the first-order corrections in 
\S\ref{firstordercorr}.
Adding the lowest-order contributions to the scalar masses from $V_F$ and $V_{\rm up}$, we obtain
\begin{align}
\partial^2_{a \bb} V_{\rm tot} \axs &=  K^{c \bar{d}} m_{ac} m_{\bb \bar{d}} + \frac{4}{3} \frac{\mgrs}{\Ms} K_a K_{\bb}\,, \\
\partial^2_{a b} V_{\rm tot} \axs &= \mgrs \left[K_{ab} + \frac{4}{3 \Ms} K_a K_b - \frac{\Ms W_{ab}}{W} \right]. \label{AppBTerm}
\end{align}
Equation  (\ref{AppBTerm}) gives the lowest-order contribution to the $B$ terms, while the  mass splittings between the scalars and fermions are given by
\be
M^{2}_{a \bar{b}} \Bigr|_{Z_{*}} = \frac{4}{3} \frac{\mgrs}{\Ms} K_a K_{\bb}\,,
\ee
to this order in perturbation theory.

\subsection{Vacua for a D3-brane on the conifold}
We remind the reader that the four-dimensional effective theory we are studying is given by
\begin{align}
K &= - 3 \Ms \ln\   U = -3 \Ms \ln(T + \bar{T} - \gamma k)\,,  \label{K} \\
W &= W_0 + W_{\rm np} = W_0 + {\cal{A}}_0 e^{- \xi}\,. \label{AppW}
\end{align}
Here $\xi = a T+ \zeta$, $\zeta = - \frac{1}{n} \ln f(z)$,   and $k = r^2$. The number 
$n$ of D7-branes determines $a=2 \pi /n$,  and $f$\ is the dimensionless embedding function of the four-cycle responsible for the nonperturbative superpotential. Since we will
need to be careful about the dimensions, it is worth mentioning that the volume modulus $T$\ is dimensionless, and $\gamma$\ has mass dimension $-2$. One can introduce fields with canonical dimension $1$, e.g. $Z^{T} = \lambda T$, $Z^i = \sigma_i z_i$ where dimensionful constants $[\lambda] = 1, [\sigma_i] = -1/2$ have been introduced. These constants are fixed by the kinetic terms, by requiring canonically normalized fields at the supersymmetric minimum,
as discussed in \S\ref{AppCanNorm}.

\subsubsection{ Supersymmetric AdS solution \label{AppSol}}
As discussed in the bulk of the paper, we find supersymmetric AdS vacua by solving the F-term equations. The equation (\ref{susyrho}) for the volume modulus
can be written as
\be
W = - \frac{a U}{3}  W_{\rm np}\,,  \label{WWnp}
\ee
which, upon
defining $\xi = a T+ \zeta$, with
$\zeta = - \frac{1}{n} \ln f(z)$, 
leads to an algebraic, transcendental equation for $U$
\be
\left( 1 + \frac{a U}{3} \right)^2 e^{-aU - a \gamma k - \xi - \bar{\xi}} = \left|\frac{W_0}{{\cal{A}}_0}\right|^2. \label{rho}
\ee
Equation (\ref{susyrho}) also determines the axionic, imaginary part of $T$,
\be
\Im (T) = - \frac{1}{a} \arg\left(\frac{- W_0 e^{\zeta}}{{\cal{A}}_0}\right). \label{axion}
\ee
In a chart where we use $z_2, z_3$ and $z_4$ as independent complex coordinates on the conifold and for the Kuperstein embedding $f(z) = \frac{z_2 - \mu}{\mu}$,  equation (\ref{D3Susy}) becomes
  \begin{align}
 - \frac{1}{n ( z_2 - \mu)} + \frac{2 a \gamma}{3 r} \left( \bar{z}_2 - \frac{\bar{z}_1 z_2}{z_1} \right) &= 0\,, \label{F2} \\
 \frac{2 a \gamma}{3 r} \left( \bar{z}_3 - \frac{ \bar{z}_1 z_3}{z_1} \right) &= 0 \label{F3}\,, \\
 \frac{2 a \gamma}{3 r} \left(\bar{z}_4 - \frac{ \bar{z}_1 z_4}{z_1} \right) &= 0\,, \label{F4}
 \end{align}
after using that $k = r^2$ far from the tip of the conifold, and that $\partial_i z_1 = - \frac{z_i}{z_1}$ for $i = 2, 3, 4$ in this chart. The radius is related to the standard complex coordinates on the conifold by $\sum^4_{a=1} |z_a|^2 = r^3$. Writing $z_A = |z_A| e^{i \eta_A}$, equations (\ref{F3}, \ref{F4}) imply that $ \eta_1 = \eta_3 = \eta_4$, but they do not restrict the norms of $z_3, z_4$. Equation (\ref{F2}) on the other hand can be written as
\be
 \frac{1}{\left(|z_2| \pm |\mu| \right)^3} = 4 \left(\frac{4 \pi \gamma}{3} \right)^3  |z_2|\,.
 \ee
The different  signs come from choosing either $\eta_2 = \eta_{\mu} $ for the upper sign or  $\eta_2 = \eta_{\mu} + \pi$ for the lower, reflecting the fact that there are two distinct supersymmetric loci for the D3-brane: one located just \emph{above} the D7-branes, and the other located far down the throat. Since the norms of $z_3$ and $z_4$ are undetermined, we have a two-dimensional moduli space. We choose to do our analysis for the point where $z_3 = z_4 = 0$ and $z_2$ is real by choice of the phase of $\mu$.

\subsubsection{Other D7-brane embeddings}
As an illustration of the fact that our supersymmetric solutions are generic for a large class of D7-brane embeddings, we derive the kindred solution for the Ouyang embedding \cite{Ouyang} specified by the embedding function $f(w) = \frac{w_2 - \mu}{\mu}$ when written in terms of the $w$-coordinates of equation (\ref{Wcoords}).
The solution has a moduli space consisting of two isolated points satisfying the equation
\be
\frac{\omega}{\omega \pm \mu} = \frac{2 an \gamma}{3} \omega^{4/3}\,,
\ee
One of these solutions is located at $\omega \gtrsim \mu$, while the other is at $\omega \gtrsim 0$. In this notation $w_2 = \omega e^{i \eta}$, where the phase $\eta$\ is fixed to be the phase of $\mu$ --- possibly up to a phase difference $\pi$, and here $r = \omega^{2/3}$. This illustrates that supersymmetric solutions are generic, and that the pairing of solutions that we have commented upon may be a feature of a wide variety of D7-brane embeddings.

\subsection{Canonical Normalization \label{AppCanNorm}}
In order to correctly assess the scaling of the masses, we obtain the canonically normalized fields. With the Kähler potential (\ref{K}), the Kähler metric is given by
\be
K_{a \bar{b}} = 3 \Ms \left[ \frac{U_a U_{\bar{b}}}{U^2} - \frac{U_{a \bar{b}}}{U} \right].
\ee
We find that, in our case, the diagonalization of this metric is essentially captured by choosing the constants $\lambda, \sigma_i$ appropriately:  
\begin{align}
\lambda &:= \left( \frac{3 \Ms}{U^2} \right)^{1/2}\,,  \\
\sigma_2 &:= \left( \frac{ \Ms}{ U} \left[ \frac{B}{ \pi \mu_0^2 } \right] \right)^{1/2}\,, \\
\sigma_3 &:= \left( \frac{3 \Ms}{ U} \left[ \frac{B}{4 \pi \mu_0^2 } \right] \right)^{1/2}\,.
\end{align}
This normalization gives
$
K_{T \bar{2}} 
=  - \frac{ \sqrt{3} }{ 2 (\pi   U   B)^{1/2} }$.
Thus, this entry is suppressed
by ${\cal{O}}(\frac{1}{\sqrt{UB}})$ with respect to the other entries in the metric, and does not affect the determinant of the metric to the order that we are working. A completely diagonal metric may be chosen by performing a unitary transformation after specifying the constants $\lambda, \sigma$. However, encouraged by the relative smallness of the off-diagonal metric elements --- for the values of $U$ and $B$ discussed in \S\ref{BoundBSect}, the off-diagonal metric elements are of order $10^{-3}$ --- we will not perform this unitary transformation that would mix the hidden sector Kähler modulus with our proxy D3-brane visible fields. To be explicit, taking two derivatives on the Kähler potential and evaluating it at the supersymmetric point with this definition of $\lambda$ and $\sigma_i$ for the canonically normalized fields gives
\begin{align}
K_{MN} = \delta^{T \bar{T}, z_2 \bar{z_2}, z_3 \bar{z_3}, z_4 \bar{z_4} }_{MN} - \frac{\sqrt{3}}{2(\pi U B)^{1/2}} \left( \delta^{T \bar{2}}_{\{MN\} } + \delta^{\bar{T} 2}_{\{MN\} } \right) + \\
+ \left[  \delta_{MN}^{T T, 33, 44} - \frac{1}{2} \delta^{22}_{MN} - \frac{\sqrt{3}}{2(\pi U B)^{1/2}}  \delta^{T 2}_{\{MN\} } + c.c. \right]\,.  \label{Kmn}
\end{align}
Here, a $\delta$-function with two indices is a shorthand for two delta functions. The curly braces correspond to symmetrization, without a factor of $\frac{1}{2}$, i.e. $\delta^{PQ}_{\{MN\}} =\delta^P_M \delta^Q_N + \delta^P_N \delta^Q_M$. As usual, only the holomorphic+antiholomorphic derivatives correspond to metric elements.
The inverse Kähler metric is 
given, to leading order in $1/B$ and to second order in $1/(aU)$, by
$
K^{a \bar{b}} = \delta^{a \bar{b}} + \frac{\sqrt{3}}{2(\pi U B)^{1/2}}  \delta^{a \bar{b}}_{T \bar{2}, 2 \bar{T}}.
$
It is well-known and  easy to verify that the Kähler metric is no-scale,
$\KAB K_A K_{\bar{B}} = 3 \Ms$, and that
$\KAB K_A = - \lambda U\  \delta^{\bar{B}}_{\bar{T}}$,
by using
\be
K_M = -\sqrt{3} \M  \delta_M^{T} + \frac{3 \M}{2 (\pi U B)^{1/2}} \delta_M^2 + c.c. \label{Km}
\ee

\subsection{Details of the mass matrix \label{AppMass}}
Using equation \eqref{Kmn} and the fact that
\be
\partial_a D_b W = - \frac{W}{\Ms} \left(  (aU +2) \delta_{ab}^{T T} + \frac{1}{2} \delta_{ab}^{2 2} - \delta_{ab}^{33, 44} - \frac{\sqrt{3}(aU +2) }{2 (\pi U B)^{1/2}} \delta_{\{a b \} }^{T 2}  \label{daFb}
\right) \, ,
\ee
the AdS supersymmetric masses  \eqref{Fmassabb} and \eqref{Fmassab}
are easily evaluated.
We find that in our case, the AdS supersymmetric $B$ terms, denoted ${\cal B}_{ab}$ to distinguish
them from the Minkowski space $B$ term, turn out to be real.
Thus, the mass matrix separates into two blocks
when written in terms of real fields, $Z^a = X^a + i Y^a$,
as
$
V_F
\supset ({\cal M}^2_{\rm tot})_{a \bar{b}} Z^a \bar{Z}^{\bar{b}} + \frac{1}{2}( {\cal B}_{ab} Z^a Z^b + h.c.)
= {\cal M}^2_{X^a X^b} X^a X^b + { \cal M}^2_{Y^a Y^b} Y^a Y^b,
$ with
\begin{align}
 {\cal M}^2_{X^a X^b} =&  \left(  ({\cal M}^2_{\rm tot})_{a \bar{b}} + {\cal B}_{ab}  \right), \\
 { \cal M}^2_{Y^a Y^b} =&  \left( ({\cal M}^2_{\rm tot})_{a \bar{b}} - {\cal B}_{ab} \right).
\end{align}
Here ${\cal B}_{ab} = \partial_a \partial_{b} V_F |_{Z_{0}} $, and $({\cal M}^2_{\rm tot})_{a \bar{b}} = \partial_{a \bb} V_F |_{Z_{0}}  $ denotes the total AdS scalar mass. After the canonical normalization discussed in \S\ref{AppCanNorm}, the resulting scalar mass matrices  are most transparently written in terms of  real fields for which we have --- to leading order in $1/B$ and to second order in $1/(aU)$ --- the supersymmetric masses:
\be
\partial^2_{M N} V_F\  \axs = \nonumber
\ee
\be
&=&  \left(
\begin{array}{c c c c c c c c}
 { \scriptstyle m^2_{T \bar{T}} +  m^2_{T T} } & {\scriptstyle m^2_{T \bar{2}} +  m^2_{T 2} } \\
 { \scriptstyle  m^2_{T \bar{2}} +  m^2_{T 2} } &  { \scriptstyle  m^2_{2 \bar{2}} +  m^2_{2 2} } & \\
 & &  { \scriptstyle m^2_{3 \bar{3}} +  m^2_{3 3 } }& \\
& & &  { \scriptstyle m^2_{4 \bar{4}} +  m^2_{4 4 }} & \\
 & & & & { \scriptstyle  m^2_{T \bar{T}} -  m^2_{T T} }&  { \scriptstyle  m^2_{T \bar{2}} -  m^2_{T 2} } \\
 & & & &   { \scriptstyle m^2_{T \bar{2}} -  m^2_{T 2}} & { \scriptstyle m^2_{2 \bar{2}} -  m^2_{2 2}} & \\
 & & & &  & &  { \scriptstyle m^2_{3 \bar{3}} -  m^2_{3 3 } }& \\
 & & & & & & &  { \scriptstyle  m^2_{4 \bar{4}} -  m^2_{4 4 }}
\end{array} \right) \nonumber
\\
\ee
\be
&=& \frac{|V_F|}{2 \Ms} \left(
\begin{array}{c c c c c c c c}
 { \scriptstyle \frac{2}{3}(a^2 U^2 + 5aU) } &   { \scriptstyle - \frac{a^2 U^2 + 5 aU}{\sqrt{3} (\pi UB)^{\frac{1}{2}}} }\\
 { \scriptstyle - \frac{a^2 U^2 + 5 aU}{\sqrt{3} (\pi UB)^{\frac{1}{2}}} } &  { \scriptstyle \frac{a^2 U}{2 \pi B} - \frac{5}{6}} & \\
 & &  { \scriptstyle -\frac{4}{3} } & \\
& & &  { \scriptstyle -\frac{4}{3} } & \\
& & & &  { \scriptstyle \frac{2}{3}(a^2 U^2 + 3aU)}  &  { \scriptstyle - \frac{a^2 U^2 + 3 aU}{\sqrt{3} (\pi UB)^{\frac{1}{2}}} }\\
& & & &  { \scriptstyle  - \frac{a^2 U^2 + 3 aU}{\sqrt{3} (\pi UB)^{\frac{1}{2}}} } &  { \scriptstyle \frac{a^2 U}{2 \pi B} - \frac{3}{2} } & \\
%
%
 & & & &  & &  { \scriptstyle 0} & \\
 & & & & & & &   { \scriptstyle 0}
\end{array} \right)\,. \label{SusyMassMatrix}
\ee
Here $ V_F = V_F\axs = \Lambda_{\rm AdS} \Ms = - 3 \mgrs \Ms$. From this expression, it is evident that there are several  tachyonic directions at this AdS vacuum. The stability of the solution requires masses larger than the  Breitenlohner-Freedman mass, which in $AdS_4$ is ${\cal M}^2_{\rm BF} = - \frac{3}{2} \frac{|V_F|}{\Ms}$. The eigenvalues of matrix (\ref{SusyMassMatrix}) are $\frac{|V_F|}{\Ms} \left[ \frac{1}{3}(a^2 U^2 + 5aU), -\frac{5}{12}, -\frac{2}{3}, -\frac{2}{3},  \frac{1}{3}(a^2 U^2 + 3aU), -\frac{3}{4}, 0, 0    \right]$, to leading order in $1/B$ and to second order in $1/aU$,
so as expected there is no instability. The supersymmetric mass-splittings can be read off from equation (\ref{MassSplit}),
\be
{\cal M}^{2\ }_{a \bar{b}} \Bigr|_{Z_{0}} = - 2 e^{K/ \Ms}  \frac{|W|^2}{M^4_{Pl}} K_{a \bb} = - 2\  \mgrs K_{a \bb}
\ee
at the AdS supersymmetric point, while the $B$ term masses are proportional to the fermion masses,
\begin{align}
{\cal B}_{ab} |_{Z_{0}} &= \partial^2_{ab} V_F \Bigr|_{Z_{0}} = - \frac{e^{K/\Ms} \overline{W}}{ \Ms} \partial_{ a } D_{ b }W  \\
&= \frac{|V_F|}{3 \Ms} \left\{ (aU + 2) \delta_{ab}^{T T} + \frac{1}{2} \delta_{ab}^{2 2} - \delta_{ab}^{33, 44} - \frac{\sqrt{3} (aU + 2)}{2 (\pi U B)^{1/2}} \delta_{\{ ab\}}^{T 2} \right\}.
\end{align}

\subsubsection{Mass matrix after uplift \label{AppDeltaM}}
After incorporating the supersymmetry breaking by adding the uplift potential to the F-term potential, the vacuum expectation values of the moduli get slightly modified. In this section we confirm that this shift in vevs is indeed small, and we demonstrate the surprising fact that the D3-brane does not move upon uplifting to this order in perturbation theory. To compute the shift we need the inverse of the total mass matrix at the supersymmetric point,
\footnote{This formula applies to all coordinates except $Y^3$ and $Y^4$. In these directions there is no shift to this order by the vanishing of the mass matrix and first derivative on the uplift potential.}
\be
\delta Z^M = -   \left( \partial_N \partial_M ( V_F + V_{\rm up} ) \right)^{-1}  \axs \partial_N V_{up} \axs. \label{Shift}
\ee
It is easy to see that $\partial_N V_{\rm up} = \frac{2}{3} \frac{|V_F|}{\Ms} K_N = 2 \mgrs K_N$, while two derivatives on the uplift potential can be written as in equation (\ref{mup}). Together with the contribution from the F-term potential and expressed in the real basis, the full mass matrix at $Z_0$ is
\be
\partial^2_{MN} \left( V_{\rm tot} \right) \axs = \nonumber
\ee
\be
= \frac{|V_F|}{2 \Ms} \left(
\begin{array}{c c c c c c c c}
 { \scriptstyle \frac{2}{3}(a^2 U^2 + 5aU)}  &  { \scriptstyle - \frac{a^2 U^2 + 5 aU}{\sqrt{3} (\pi UB)^{\frac{1}{2}}} } \\
 { \scriptstyle - \frac{a^2 U^2 + 5 aU}{\sqrt{3} (\pi UB)^{\frac{1}{2}}} } &  { \scriptstyle \frac{a^2 U}{2 \pi B} - \frac{1}{6} }& \\
 & &  { \scriptstyle +\frac{4}{3} }& \\
& & &  { \scriptstyle +\frac{4}{3} }& \\
& & & &  { \scriptstyle \frac{2}{3}(a^2 U^2 + 3aU) } &   { \scriptstyle- \frac{a^2 U^2 + 3 aU}{\sqrt{3} (\pi UB)^{\frac{1}{2}}} }\\
& & & &  { \scriptstyle - \frac{a^2 U^2 + 3 aU}{\sqrt{3} (\pi UB)^{\frac{1}{2}}}} &  { \scriptstyle \frac{a^2 U}{2 \pi B} + \frac{1}{2} }& \\
%
%
 & & & &  & &  { \scriptstyle 0} & \\
 & & & & & & &  { \scriptstyle 0}
\end{array} \right)\,.\,\,
\ee
Inverting, we find that to this order the only shift is in the real part of $T$ and is given by   $\frac{1}{2}( \delta Z^{T} + \delta \bar{Z}^{\bar{T}}) = \frac{\sqrt{3} \M}{aU(aU+5)}$.

To  order $1/(aU)^2, 1/B$ the fermion mass matrix at $Z_0$ is obtained by evaluating (\ref{FermAtX0}),
\be
m_{ab} \Bigr|_{Z_{0}} = e^{K/(2 \Ms)} \frac{W}{\Ms} \left\{ -(aU + 2) \delta_{ab}^{T T} - \frac{1}{2} \delta^{22}_{ab} + \delta_{ab}^{33,44} + (aU + 2) \frac{\sqrt{3}}{2 (\pi U B)^{1/2} } \delta_{\{ ab\}}^{T 2} \right\}.~\,\,\, 
\ee
The prefactor $e^{K/2\Ms}  \frac{W}{\Ms} $ is just $ m_{3/2} \axs$.
The gravitino mass is unchanged to this order from
its
value in AdS,
\be
m^2_{3/2} = \frac{|V_F| |_{Z_{0}} }{3 \Ms}\,. 
\ee
The mass splittings between the scalars and fermions in the Minkowski solution are, to this order,
\begin{align}
 M^{2}_{a \bar{b}} |_{Z_{\star}}   &= {\cal M}^{2}_{a \bar{b}} |_{Z_{0}}   + \partial^2_{a \bb} V_{\rm up} |_{Z_{0}}  = \\
 &= 4 \mgrs
 \left(
 \begin{array}{c c c c}
 1 & - \frac{\sqrt{3}}{2 (\pi U B)^{1/2} }  \\
- \frac{\sqrt{3}}{2 (\pi U B)^{1/2} }   & \frac{3}{4 \pi U B} \\
  & & 0 \\
  & & & 0
 \end{array} 
 \right)\,. 
\end{align}
The $B$ terms are, to this order,
\begin{align}
B_{ab} \equiv \partial_{ab} V_{\rm tot} |_{Z_{\star}}   &= {\cal B}_{ab} |_{Z_{0}}  + \partial_{ab} V_{\rm up} |_{Z_{0}}   = \\
&= \mgrs
\left(
\begin{array}{c c c c}
aU & -  \frac{\sqrt{3  }a }{2}(\frac{U }{\pi B})^{1/2} \\
-  \frac{\sqrt{3  }a }{2}(\frac{U }{\pi B})^{1/2} & -\frac{1}{2} \\
& & 1 \\
& & & 1 \\
\end{array} 
\right)\,. 
\end{align}

\subsubsection{Corrections to the mass matrix due to the shift in the volume modulus}
\label{firstordercorr}
We determine the relative importance of the different contributions by computing the 
first-order correction to the mass matrix in perturbation theory.  The contribution from the uplift potential comes from evaluation of equation (\ref{UpMassCorr}).
Three derivatives on the Kähler potential,
evaluated at $Z_0$, can be written as
\begin{align}
K_{MNP} |_{Z_{0}}   = \frac{1}{\M} \left\{  (\pi U B)^{1/2} \left( \frac{2}{9} \dt^{2 2 2}  + \frac{2}{3} \dt^{33\bar{2}, 44 \bar{2}}  - \frac{4}{3} \dt^{332, 442} - \frac{1}{3} \dt^{22\bar{2}} - \frac{1}{3} \dts^{3\bar{3}2, 4 \bar{4} 2}                 \right)      -  \right. \nonumber \\
\left.  -  \frac{1}{\sqrt{3}} \left( \dts^{2\bar{2} T, 3\bar{3} T, 4\bar{4} T} + \frac{2}{3} \dt^{T T T}  + 2 \dt^{T T \bar{T}} - \frac{1}{2} \dt^{22T} - \frac{1}{2} \dt^{22 \bar{T}}  + \dt^{33 T, 44 T} + \dt^{33 \bar{T}, 44 \bar{T} }\right) + \right.   \nonumber \\
\left. + \frac{1}{2 (\pi U B)^{1/2}} \left[ \frac{5}{2} \dt^{T T 2} + 2 \dt^{T T \bar{2}} + 2 \dts^{T \bar{T} 2}  \right]  + c.c. +  {\rm perm }_{(M, N) \leftrightarrow P} \right\} \, . \label{Kmnp}
\end{align}
After taking ${\rm perm }_{(M,N) \leftrightarrow P} $ into account we read off that e.g.
$K_{222} = \frac{2 \sqrt{\pi U B}}{3 \M}$.
Remembering that the shift is only in the direction of the volume modulus, we can immediately estimate the size of the contributions to the mass matrix from equation (\ref{UpMassCorr}). Recall from equations (\ref{Kmn}, \ref{Km}) that $K_{MN}$ is no larger than ${\cal O}(1)$ and that $K_M$ is no larger than ${\cal O}(1)$. With one index being canonically normalized $T$ or $\bar{T}$, equation (\ref{Kmnp}) gives that $K_{MNP}$  is no larger than ${\cal O}(1)$. All together, $\partial^3 V_{\rm up}$ is no larger than ${\cal O}( \frac{|V_F|}{\M^3})$, but the shift in the real part of the Kähler modulus scales like $\delta X^{T} \sim \frac{\M}{(aU)^2}$. The contribution to the mass matrix from the uplift potential will therefore scale like $\delta X^{T} \ \partial^3 V_{\rm up} \axs \sim \frac{|V_F|}{\Ms} \frac{1}{(aU)^2}$. The smallest non-vanishing entry in $\partial^2 V_{\rm tot} \axs \sim 1$, so we conclude that the first-order correction from the uplift potential will come in  at a
subleading order in $1/(aU)$ and can consistently be dropped. By direct evaluation we find that $\delta m^2_{3 \bar{3}} = \delta B_{33}$, which means that the flat directions are not lifted by the uplift potential.

The first-order correction from the F-term potential is more tedious to obtain, but follows from straightforward evaluation of equation (\ref{Vabc}) and the corresponding equation for all holomorphic indices. These terms also do not contribute  before order $1/aU$.

Finally, the magnitude of the F-terms for the canonically normalized fields  can be found to linear order by contracting equation \eqref{daFb} with the shift \eqref{Shift}.

\subsection{A bound on B \label{AppBoundB}}
Requiring that both the D3-brane and D7-brane are located in the warped region gives a bound on $B$. 
Since we are considering only solutions in which the D3-brane lies deeper down in the throat than the D7-brane, $B$ is bounded from below: 
$B > 1$. The upper bound comes from considering the arguments in \cite{Baumann:2006cd} and \cite{BDKM}, in which a bound on
$k$ is obtained
in terms of $N$, the number of D3-branes that make up the throat before we
add our toy visible sector:
\be
\frac{\gamma k}{T + \bar{T}} \leq \frac{2}{3} \frac{1}{N}\,. \label{bound}
\ee
For the D7-brane to extend down the throat,  the bound (\ref{bound}) should apply if evaluated at the point of lowest descent of the D7-brane into the throat.
We have $\gamma k_{\rm D7} = \frac{3}{4 \pi} B^{1/3}$, which follows  from direct evaluation of $B$.
It follows that
\be
B \leq \left(\frac{8 \pi}{9}\right)^3 \left(\frac{T + \bar{T}}{N}\right)^3\,. 
\ee
In this case $U = T + \bar{T} - \gamma k_{\rm D7} = T + \bar{T} - \frac{3}{4 \pi} B^{1/3}$. If $ 4^3 U^3 \gg B$, then the above bound can be written as
\be
B \leq L^3\,.
\ee
Together these bounds imply that $1 < B \leq L^3$.
In particular, a consistent solution requires that  $L >1 $, from which it follows that
\be
L = \frac{8 \pi}{9} \frac{U}{N} \approx 2.79 \frac{U}{N}  > 1\,.
\ee

\end{document}